\documentclass[10pt,twocolumn,twoside]{IEEEtran}

\usepackage{graphicx}
\usepackage{amsmath}
\usepackage{float}
\usepackage{amssymb}
\usepackage{latexsym}
\usepackage{xfrac}
\usepackage{subcaption}
\usepackage{amsfonts}
\usepackage{amstext}
\usepackage{amsthm}
\usepackage{dsfont}
\usepackage[normalem]{ulem}
\usepackage{latexsym}
\usepackage{soul}
\usepackage{booktabs}
\usepackage{amsfonts,amssymb,amsmath}
\usepackage{hyperref}
\usepackage{color}
\usepackage[dvipsnames]{xcolor}
\usepackage{xspace}
\usepackage[ruled,vlined]{algorithm2e}

\usepackage{array}
\newcolumntype{P}[1]{>{\centering\arraybackslash}p{#1}}
\def\BibTeX{{\rm B\kern-.05em{\sc i\kern-.025em b}\kern-.08em T\kern-.1667em\lower.7ex\hbox{E}\kern-.125emX}}
\colorlet{mygreen}{green!60!gray}

\DeclareMathOperator*{\argmax}{arg\,max}
\DeclareMathOperator*{\argmin}{arg\,min}

\begin{document}

\title{An Overview of Backdoor Attacks Against Deep Neural Networks and Possible Defences}

\author{Wei Guo, Benedetta~Tondi, \IEEEmembership{Member, IEEE}, Mauro~Barni, \IEEEmembership{Fellow, IEEE} %
\thanks{The authors are with the Department of  Information Engineering and Mathematics, University of Siena, 53100 Siena, ITALY. Wei Guo was supported by China Scholarship Council under No.201908130181. Indicate corresponding author Wei Guo.}}

\maketitle

\begin{abstract}

Together with impressive advances touching every aspect of our society, AI technology based on Deep Neural Networks (DNN) is bringing increasing  security concerns. While attacks operating at test time have monopolised the initial attention of researchers, backdoor attacks, exploiting the possibility of corrupting DNN models by interfering with the training process, represents a further serious threat undermining the dependability of AI techniques. In a backdoor attack, the attacker corrupts the training data so to induce an erroneous behaviour at test time. Test time errors, however, are activated only in the presence of a triggering event corresponding to a properly crafted input sample. In this way, the corrupted network continues to work as expected for regular inputs, and the malicious behaviour occurs only when the attacker decides to activate the backdoor hidden within the network. In the last few years, backdoor attacks have been the subject of an intense research activity focusing on both the development of new classes of attacks, and the proposal of possible countermeasures. The goal of this overview paper is to review the works published until now, classifying the different types of attacks and defences proposed so far. The classification guiding the analysis is based on the amount of control that the attacker has on the training process, and the capability of the defender to verify the integrity of the data used for training, and to monitor the operations of the DNN at training and test time. As such, the proposed analysis is particularly suited to highlight the strengths and weaknesses of both attacks and defences with reference to the application scenarios they are operating in.
\end{abstract}

\begin{IEEEkeywords}

Backdoor attacks, backdoor defences, AI Security, Deep Learning, Deep Neural Networks
\end{IEEEkeywords}

%

\section{INTRODUCTION}

Artificial Intelligence (AI) techniques based on Deep Neural Networks (DNN) are revolutionising the way we process and analyse data, due to their superior capabilities to extract relevant information from complex data, like images or videos, for which precise statistical models do not exist. On the negative side, increasing concerns are being raised regarding the security of DNN architectures when they are forced to operate in an adversarial environment, wherein the presence of an adversary aiming at making the system fail can not be ruled out. In addition to attacks operating at test time, with an increasingly amount of works dedicated to the development of suitable countermeasures against adversarial examples \cite{szegedy2013intriguing,goodfellow2014explaining}, attacks carried out at training time have recently attracted the interest of researchers \cite{biggio2012poisoning, gonz17, gu_badnets_2017,chen_targeted_2017}. Among them, {\em backdoor attacks} are raising increasing concerns due to the possibility of stealthily injecting a malevolent behaviour within a DNN model by interfering with the training phase. The malevolent behaviour (e.g., a classification error), however, occurs only in the presence of a triggering event corresponding to a properly crafted input. In this way, the {\em backdoored} network continues working as expected for regular inputs, and the malicious behaviour is activated only when the attacker feeds the network with a triggering input.

The earliest works demonstrating the possibility of injecting  a backdoor into a DNN have been published in 2017 \cite{gu_badnets_2017, chen_targeted_2017, ji_backdoor_2017,liu2017neural}. Since then, an increasing number of works have been dedicated to such a subject, significantly enlarging the class of available attacks, and the application scenarios potentially targeted by backdooring attempts.
The proposed attacks differ on the basis of the event triggering the backdoor at test time, the malicious behaviour induced by the activation of the backdoor, the stealthiness of the procedure used to inject the backdoor, the  modality through which the attacker interferes with the training process, and the knowledge that the attacker has about the attacked network.

As a reaction to the new threats posed by backdoor attacks, researchers have started proposing suitable solutions to mitigate the risk that the dependability of a DNN is undermined by the presence of a hidden backdoor. In addition to methods  to reveal the presence of a backdoor, a number of solutions to remove the backdoor from a trained model have also been proposed, with the aim of producing a cleaned model that can be used in place of the infected one \cite{liu_fine-pruning_2018,wang_neural_2019,guo_tabor_2019}. Roughly speaking, the proposed solutions for backdoor detection can be split into two categories: methods detecting the backdoor injection attempts at training time, e.g. \cite{tran_spectral_2018,chen_detecting_2019}, and methods detecting the presence of a backdoor at test time, e.g., \cite{chen_detecting_2019, chou2020sentinet, gao_strip_2019,chen_deepinspect_2019}.  Each defence targets a specific class of attacks and usually works well only under a specific threat model.

As it always happens when a new research trend appears, the flurry of works published in the early years have explored several directions with only few and scattered attempts to systematically categorise them.
Time is ripe to look at the work done until now, to classify the  attacks and defences proposed so far, highlighting their suitability to different application scenarios, and evaluate their strengths and weaknesses.
To the best of our knowledge, the only previous attempts  to survey backdoor attacks  against DNN and defences are \cite{LiuMCZJXS20survey, chen2020backdoor_survey}, with the former work having
a limited scope, and the latter which focuses on a specific attack surface, namely, the outsourced cloud environment.
An overview paper addressing all the application domains of backdoor attacks have also been published in ~\cite{li2020backdoor_survey}. With respect to such  an overview, we make the additional effort to provide a clear definition of the threat models, formalizing the requirements that attacks and defences must satisfy in the various settings. This helps us to cast all the backdoor attacks and defences developed so far under a unique umbrella.
%

To be more specific, the  contributions of the present work can be summarised as follows:
\begin{itemize}
  \item We provide a formalization of backdoor attacks, defining the possible threat models and the corresponding requirements (Section \ref{sec.threatmodels}). A rigorous description of the threat models under which the backdoor attacks and defences operate is, in fact, a necessary step for a proper security analysis. We distinguish between different scenarios  depending on the  control that the attacker has on the training process. In particular, we distinguish between i) {\em full control} attacks, wherein the attacker is the trainer herself, who, then, can interfere with every step of the training process, and ii) {\em partial control} attacks, according to which the attacker controls the training phase only partially. The requirements that
 attacks and defences must satisfy in the various settings are also described, as they are closely related to the threat models.
        \item We systematically review the backdoor attacks proposed so far, specifying the control scenario under which they can operate and their limitations (Section \ref{sec.attacks}). Specifically, we distinguish between two classes of methods: i) {\em corrupted-label attacks}, that is attacks tampering the labels of the poisoned samples, and ii) {\em clean-label attacks}, according to which the attacker can not change or define the labels of the poisoned samples.
  \item We provide a thorough review of possible defences,  by casting them in the classification framework defined previously. In particular, we categorize the defences based on the levels at which they operate, that is:  i) {\em data} level, ii) {\em model} level, and iii) {\em training dataset} level.  The defences within each category are further classified based on the approach followed for the detection and the removal of the backdoor. Thanks to the proposed classification, defence methods can be compared according to the extent by which they satisfy the requirements set by the threat model wherein they operate.
      \item We point out possible directions for future
research, reviewing the most challenging open issues.
\end{itemize}

To limit the scope and length of the paper, we focus on attacks and defences in the field of image and video classification, leaving aside other application domains, e.g., natural language processing \cite{dai2019backdoor}. We also avoid discussing the emerging field of attacks and defences in collaborative learning scenarios, like federated learning, \cite{bagdasaryan2020backdoor,bhagoji2019analyzing,xie2019dba,chen2020backdoor_federated}.
Finally, we stress that the survey is not intended to review  all the methods proposed so far, on the contrary, we describe in details only the most significant works of each attack and defense category, and provide a pointer to all the other methods we are aware of.

We expect that research on backdoor attacks and corresponding defences will continue to surge in the next years, due to the seriousness of the security threats they pose, and hope that the present overview will help researchers to focus on the most interesting and important challenges in the field.

The rest of this paper is organised as follows: in Section~\ref{sec.threatmodels}, we
formalize the backdoor attacks, by paying great attention to discuss the attack surface and the possible defence points. Then, in Section~\ref{sec.attacks}, we review the literature of backdoor attacks.
Following the categorization introduced in Section~\ref{sec.threatmodels},
the defence methods are reviewed and compared in Sections ~\ref{sec.datalevel} through ~\ref{sec.datasetlevel}, classifying them according to the level (input data, model, or training dataset levels) at which they operate.
Finally, in Section~\ref{sec.con}, we discuss the most relevant open issues and provide a roadmap for future research.

\section{Formalization, Threat Models and Requirements}
\label{sec.threatmodels}


In this section, we give a rigorous formulation of backdoor attacks and the corresponding threat models, paying particular attention to the requirements that the attack must satisfy under different models. We also introduce the basic notation used in the rest of the paper.

We will assume that the model targeted by the attack aims at solving a classification problem within a supervised learning framework. Other tasks and training strategies, such as semantic segmentation~\cite{li2021hidden} or contrastive learning~\cite{carlini2021poisoning}, can also be subject to backdoor attacks, however, to avoid expanding too much the scope of the survey, and by considering that most of existing literature focuses on classification networks, we will restrict our discussion to this kind of tasks.

\subsection{Basic Notation and Formalization}
\label{sec.notation}

In supervised learning, a classifier $\mathcal{F}_{\theta}$ is trained to map a sample $x$ from the input space $\mathbb{X}$ into a label $y$ belonging to the label space $\mathbb{Y}=\{1,...,C\}$. Classification is usually (but not necessarily) achieved by:
\begin{equation}
	\mathcal{F}_{\theta}(x)=\argmax(f_{\theta}(x)),
\end{equation}
where $f_{\theta}$ is a \textit{C}-element vector $f_{\theta}(x)$, whose elements represent the probabilities over the labels in $\mathbb{Y}$ (or some other kind of soft values), and $\argmax(\cdot)$ outputs the index with the highest probability. In the following, we indicate the $k$-th element of $f_{\theta}(x)$ as $[f_{\theta}(x)]_k$, and the output of the $i$-th layer of the network as $f^{i}_{\theta}(x)$. Here, $\theta$ indicates the trainable parameters of the model. $\mathcal{F}$ may also depend on a set of hyperparameters, denoted by $\psi$, defining the exact procedure used to train the model (e.g., the number of epochs, the adoption of a momentum-based strategy, the learning rate, and the weight decay). Unless necessary, we will not indicate explicitly the dependence of $\mathcal{F}$ on $\psi$. ${\mathcal{F}}_{\theta}$ is trained by relying on a training set $\mathcal{D}_{tr}=\{(x^{tr}_{i}, y^{tr}_{i}),i=1,...,|\mathcal{D}_{tr}|\}$, where $(x^{tr}_{i}, y^{tr}_{i})\in \mathbb{X}\times \mathbb{Y}$ and $|\mathcal{D}_{tr}|$ indicates the cardinality of $\mathcal{D}_{tr}$. The goal of the training procedure is to define the parameters $\theta$, by solving the following general optimization problem:
\begin{equation}
\label{equ:optimization}
	\argmin_\theta \sum_{i=1}^{|\mathcal{D}_{tr}|}L(f_{\theta}(x^{tr}_{i}),y^{tr}_{i}),
\end{equation}
where $L$ is a loss function closely related to the classification task the network has to solve.


At testing time, the performance of the trained model ${\mathcal{F}}_{\theta}$ are evaluated on the elements of a test dataset $\mathcal{D}_{ts}=\{(x^{ts}_{i}, y^{ts}_{i}),i=1,...,|\mathcal{D}_{ts}|\}$. In particular, the accuracy of the model is usually evaluated as follows:
\begin{equation}
	\mathcal{A}(\mathcal{F}_{\theta}, \mathcal{D}_{ts})=\frac{\#\{{\mathcal{F}}_{\theta}(x^{ts}_{i})=y^{ts}_{i}\}}{|\mathcal{D}_{ts}|},
\end{equation}
where $\#\{{\mathcal{F}}_{\theta}(x^{ts}_{i})=y^{ts}_{i}\}$ indicates the number of successful predictions.

\subsection{Formalization of Backdoor Attacks}
\label{sec.backdef}

As we briefly discussed in the Introduction, the goal of a backdoor attack is to make sure that, at test time, the backdoored model behaves as desired by the attacker in the presence of specific triggering inputs, while it continues to work as expected on normal inputs.
To do so, the attacker interferes with the generation of the training dataset. In some cases (see section \ref{subsubFC}), she can also shape the training procedure, so to directly instruct the network to implement the desired behaviour.

Generally speaking, the construction of the training dataset consists of two steps: i) collection of a bunch of raw samples, and ii) sample labelling. During the first step, the attacker injects into the training dataset a set of poisoned samples $(\tilde{x}_{1}^{tr},\tilde{x}_{2}^{tr},...)$, where each element {\em contains} a triggering pattern $\upsilon$.
The shape of the triggering pattern and the exact way the pattern is associated to the poisoned samples depends on the specific attack and it will be detailed later. Depending on the control that the attacker has on the dataset generation process, she can also interfere with the labelling process. Specifically, two kinds of attacks are possible. In a \textit{corrupted-label attack}, the attacker can directly label $\tilde{x}_{i}^{tr}$, while in a \textit{clean-label attack}, the labelling process is up to the legitimate trainer.

Let us indicate with $\tilde{y}_{i}^{tr}$, the label associated to $\tilde{x}_{i}^{tr}$. The set with the labeled poisoned samples forms the poisoning dataset $\mathcal{D}_{tr}^{p}=\{(\tilde{x}_{i}^{tr}, \tilde{y}_{i}^{tr}),i=1,...,|\mathcal{D}_{tr}^{p}|\}$. The poisoning dataset is merged with the benign dataset $\mathcal{D}_{tr}^{b}=\{(x_{i}^{tr}, y_{i}^{tr}),i=1,...,|\mathcal{D}_{tr}^{b}|\}$ to generate the poisoned training dataset $\mathcal{D}_{tr}^{\alpha}=\mathcal{D}_{tr}^{b}\cup \mathcal{D}_{tr}^{p}$, where
\begin{equation}
\label{alpha}
\alpha=\frac{|\mathcal{D}_{tr}^{p}|}{|\mathcal{D}_{tr}^{p}|+|\mathcal{D}_{tr}^{b}|},
\end{equation}
hereafter referred to as poisoning ratio, indicates the fraction of corrupted samples contained in the poisoned training dataset.

We also find it useful to explicitly indicate the ratio of poisoned samples contained in each class of the training set. Specifically, let $\mathcal{D}_{tr,k}^b$ (res. $\mathcal{D}_{tr,k}^p$), indicate the subset of samples for which $y^{tr}_{i} = k$ in the benign (res. poisoned), dataset. Then, Then $\mathcal{D}_{tr}^b = \bigcup_k \mathcal{D}_{tr,k}^b$ ($\mathcal{D}_{tr}^p = \bigcup_k \mathcal{D}_{tr,k}^p$). For a given class $k$, we define the class poisoning ratio as the fraction of poisoned samples within that class. Formally, 
 \begin{equation}
	\beta_k=\frac{|\mathcal{D}_{tr,k}^{p}|}{|\mathcal{D}_{tr,k}^{p}|+|\mathcal{D}_{tr,k}^{b}|}.
\end{equation}
In the following, when the attacker poisons only samples from one class, or when it is not necessary to indicate the class affected by the attack, the subscript $k$ is omitted.

Due to poisoning, the classifier $\mathcal{F}_{\theta}$ is trained on $\mathcal{D}_{tr}^{\alpha}$, and hence it learns the correct classification from the benign dataset $\mathcal{D}_{tr}^{b}$ and the malevolent behaviour from $\mathcal{D}_{tr}^{p}$.
By assuming that the attacker does not control the training process, training is achieved by optimizing the same loss function used to train a benign classifier, as stated in the following equation:
\begin{equation}
\label{equ:poison_optimization}
	\theta_{\alpha}=\argmin_\theta
	\bigg(\sum_{i=1}^{|\mathcal{D}_{tr}^{b}|}L(f_{\theta}(x_i^{tr}),y_i^{tr})+ \sum_{i=1}^{|\mathcal{D}_{tr}^{p}|}L(f_{\theta}(\tilde{x}_{i}^{tr}), \tilde{y}^{tr}_{i})\bigg),
\end{equation}
where, for sake of clarity, we have split the loss function into two terms, one term accounting for the benign samples and the other for the poisoned ones.
In the sequel, we denote the backdoored model resulting from the optimization in \eqref{equ:poison_optimization} by $\mathcal{F}_{\theta}^{\alpha}$.

To be effective, a backdoor attack must achieve two main goals\footnote{Other goals depend on the attack scenario as discussed in Section \ref{sec.surface}.}:
\begin{itemize}
	\item {\em Stealthiness at test time}. The backdoor attack should not impair the expected performance of the model. This means that the backdoored model ${\mathcal{F}}_{\theta}^{\alpha}$
and the benign one $\mathcal{F}_{\theta}$ should have similar performance when tested on a benign testing dataset $\mathcal{D}_{ts}^{b}$, i.e., {$\mathcal{A}(\mathcal{F}_{\theta}^{\alpha}, \mathcal{D}_{ts}^{b})\simeq \mathcal{A}(\mathcal{F}_{\theta}, \mathcal{D}_{ts}^{b})$.}

	\item {\em High attack success rate}. When the triggering pattern $\upsilon$ appears at the input of the network, the malevolent behaviour should be activated with a high probability. To measure this probability, the backdoored model $\mathcal{F}_{\theta}^{\alpha}$ is evaluated upon a poisoned test dataset $\mathcal{D}_{ts}^{p}$, with samples $\tilde{x}_{ts}$ from all the classes, with the exception of the target class $t$, containing the triggering pattern, and labelled as $\tilde{y}_{ts} = t$. The attack success rate is computed as ${ASR}(\mathcal{F}_{\theta}^{\alpha},\mathcal{D}_{ts}^{p}) = \mathcal{A}(\mathcal{F}_{\theta}^{\alpha}, \mathcal{D}_{ts}^{p})$.
\end{itemize}

A list of the symbols introduced in this section and all the other symbols used throughout the paper is given in Table~\ref{tab:NS}.
\begin{table}[h!]
\centering
\small
\caption{List of symbols}
\label{tab:NS}
\begin{tabular}{|p{1.8cm}|p{6.3cm}|} \hline
	\textbf{Notation}& \textbf{Explanation} \\ \hline
	$\upsilon$ & triggering pattern \\ \hline
	$L$ & Loss function \\ \hline
	$\psi$ & Training Hyperparameters \\ \hline
	$\mathbb{X}, \mathbb{Y}$ & Input space and label space \\ \hline
	${x}, {y}$ & Benign samples and their labels \\ \hline
	$\tilde{x}, \tilde{y}$ & Poisoned samples and their labels\\ \hline
	$x^{tr},y^{tr}$ & Training samples and corresponding labels\\ \hline
	$x^{ts},y^{ts}$ & Testing samples and corresponding labels\\ \hline
	$C$ & Number of classes \\ \hline
	$\mathcal{D}_{tr}^{b}$ & Benign training dataset \\ \hline
	$\mathcal{D}_{tr}^{p}$ & Poisoned training dataset \\ \hline
	$\mathcal{D}^{\alpha}_{tr}$ & Poisoned training dataset with poisoning ratio $\alpha$\\ \hline
	$\alpha$ & Poisoning ratio \\ \hline
	$\beta_k$ & Poisoning ratio for class $k$ \\ \hline
	$\mathcal{D}_{ts}^{b}$ & Benign test dataset held by the user to evaluate the model performance \\ \hline
	$\mathcal{D}_{ts}^{p}$ & Poisoned test dataset held by the adversary to evaluate the effectiveness of the attack \\ \hline
	$\mathcal{D}_{be}$ & Benign dataset used for backdoor detection and removal \\ \hline
	$\mathcal{F}_{\theta}(\cdot)$ & Benign mode with architecture $\mathcal{F}$ and parameters $\theta$ \\ \hline
	$\mathcal{F}_{\theta}^{\alpha}(\cdot)$ & Backdoored model trained on the poisoned training dataset $\mathcal{D}_{tr}^{\alpha}$\\ \hline
	$\mathcal{F}_{\theta_{c}}(\cdot)$ & Cleaned model after backdoor removal \\ \hline
	$\hat{\mathcal{F}}_{\theta}(\cdot)$ & Surrogate or pre-trained model of $\mathcal{F}_{\theta}$ \\ \hline
	$\mathcal{F}^{meta}_{\theta}$ & Meta classifier \\ \hline
	$f_\theta(\cdot)$ & Intermediate softmax vector \\ \hline
	$f^{i}_\theta(\cdot)$ & Output of the $i$-th layer of $\mathcal{F}_\theta(\cdot)$ \\ \hline
	$[f_\theta(\cdot)]_k$ & $k$-th element of $f_\theta(\cdot)$ \\ \hline
	$\mathcal{P}(\cdot)$ & Poisoning function generating the poisoned samples $\tilde{x}$\\ \hline

	$\mathcal{E}(\cdot)$ & Feature extraction function \\ \hline
	$\mathcal{M}(\cdot)$ & Decision making function \\ \hline
	$\mathcal{A}(\cdot)$ & Accuracy metric measured on benign data  \\ \hline
	$ASR(\cdot)$ & Attack success rate measured on poisoned data \\ \hline
	$Det(\cdot)$ & Detection function \\ \hline
	$Rem(\cdot)$ & Removal function \\ \hline
   	\end{tabular}
\end{table}


\subsection{Attack surface and defence points}
\label{sec.surface}
The threat model ruling a backdoor attack, including the attack surface and the possible defence points, depends mainly on the control that the attacker has on the training process. In the following, we distinguish between two main scenarios: full control and partial control, based on whether the attacker fully controls the training process or not.

\subsubsection{Full control}
\label{subsubFC}

\begin{figure}[b!]
	\centering
	\includegraphics[width=1\columnwidth]{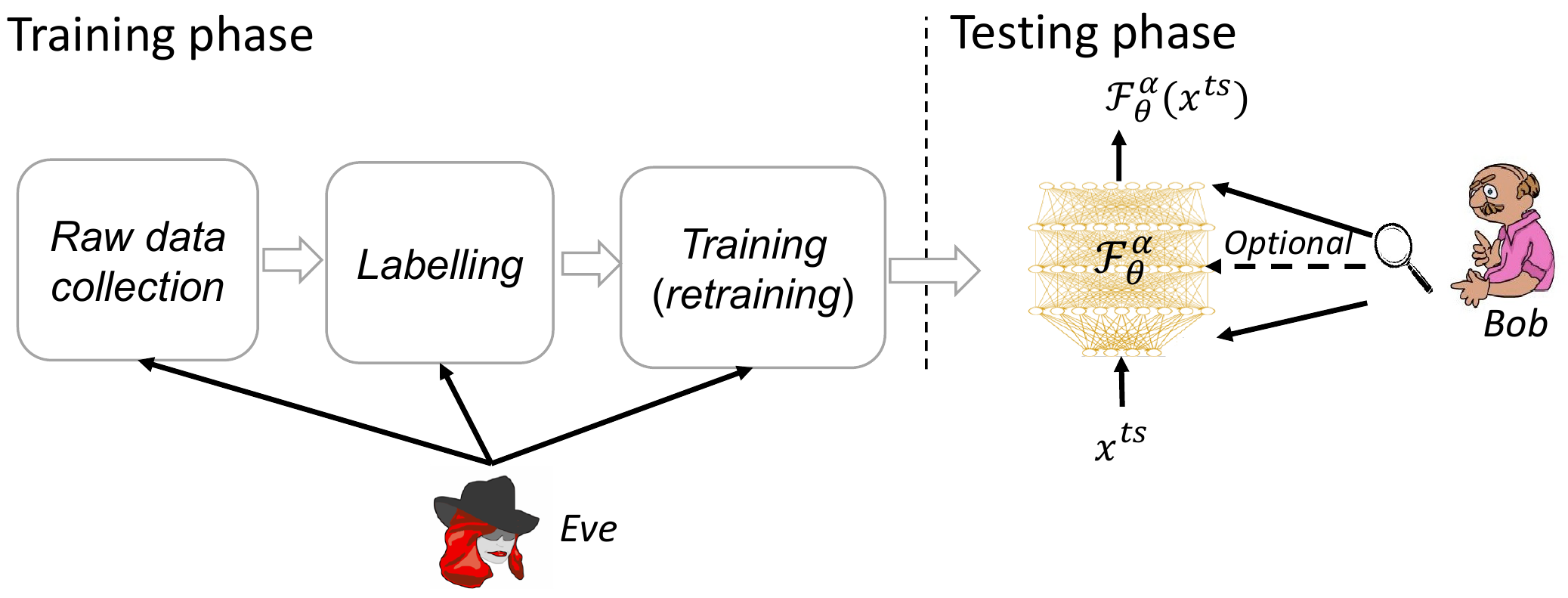}
	\caption{In the full control scenario, the attacker \textit{Eve} can intervene in all the phases of the training process, while the defender \textit{Bob} can only check the model at test time. The internal information of the model may or may not be accessible to \textit{Bob}, depending on whether the defence is a white-box or black-box one. }
	\label{fig:FC}
\end{figure}

In this scenario, exemplified in Fig.~\ref{fig:FC}, the attacker, hereafter referred to as \textit{Eve},
 is the trainer herself, who, then, can interfere with every step of the training process. This assumption is realistic in a scenario where the user, say \textit{Bob}, outsources the training task to a third-party due to lack of resources. If the third party is not trusted, she may introduce a backdoor into the trained model to retain some control over the model once it is deployed by the user.

\textbf{Attacker's knowledge and capability}: since \textit{Eve} coincides with the legitimate trainer, she knows all the details of the training process, and can modify them at will, including the training dataset, the loss function $L$, and the hyperparameters $\psi$. To inject the backdoor into the model \textit{Eve} can:
\begin{itemize}
	\item Poison the training data: \textit{Eve} designs a poisoning function $\mathcal{P}(\cdot)$ to generate the poisoned samples $(\tilde{x}_{1}^{tr},\tilde{x}_{2}^{tr},...)$ and merges them with the benign dataset.
	\item Tamper the labels: the labelling process is also ruled by \textit{Eve}, so she can mislabel the poisoned samples $\tilde{x}_{i}^{tr}$ to any class $\tilde{y}_{i}^{tr}$.
	\item Shape the training process: \textit{Eve} can choose a suitable algorithm or learning hyperparameters to solve the training optimization problem. She can even adopt an ad-hoc loss function explicitly thought to ease the injection of the backdoor~\cite{kirkpatrick_overcoming_2017}.
\end{itemize}
Other less common scenarios, not considered in this paper, may assign to the attacker additional capabilities. In some works, for instance, the attacker may change directly the weights after the training process has been completed \cite{dumford_backdooring_2018, costales_live_2020}.

\textbf{Defender's knowledge and capability}: as shown in Fig.~\ref{fig:FC}, in the full control scenario, the defender \textit{Bob} corresponds to the final user of the model, and hence he can only act at test time. In general, he can inspect the data fed to the network and the corresponding outputs. He may also query the network with untainted samples from a benign testset $\mathcal{D}^{b}_{ts}$, which is used to validate the accuracy of the network. Moreover, \textit{Bob} may hold another benign dataset $\mathcal{D}_{be}$ to aid backdoor detection or removal. In some cases, \textit{Bob} may have full access to the model, including the internal weights and the activation values of the neurons. In the following, we refer to these cases as \textit{white-box} defences. In other cases, referred to as black-box defences, \textit{Bob} can only observe the input and output values of the model.

\begin{figure}[b!]

\centering
\vspace{-0.2cm}
	\begin{subfigure}[b]{.5\textwidth}
  		\centering
  		\includegraphics[width=0.9\linewidth]{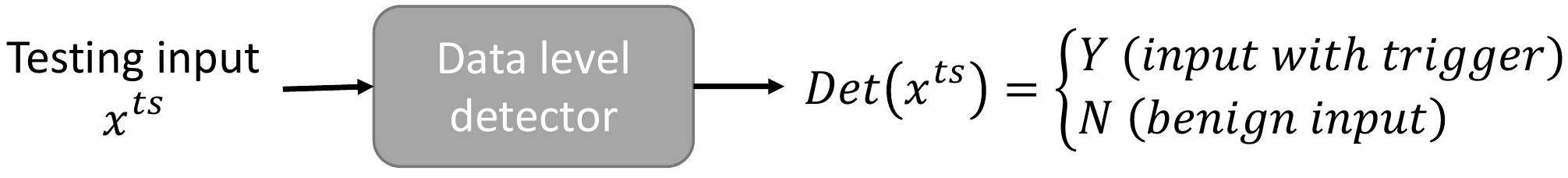}
  		\caption{Data level}
  	\label{FIG:detectionI}
	\end{subfigure}
	\\~\\
	\begin{subfigure}[b]{.5\textwidth}
  		\centering
  		\includegraphics[width=0.9\linewidth]{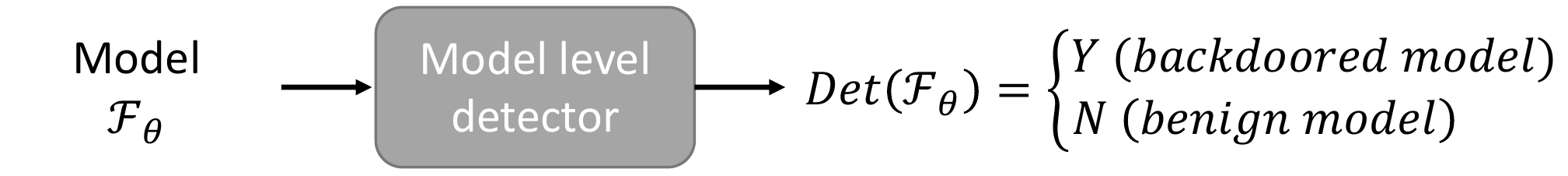}
  		\caption{Model level}
  	\label{FIG:detectionM}
	\end{subfigure}
	\\~\\
	\begin{subfigure}[b]{.5\textwidth}
  		\centering
  		\includegraphics[width=0.9\linewidth]{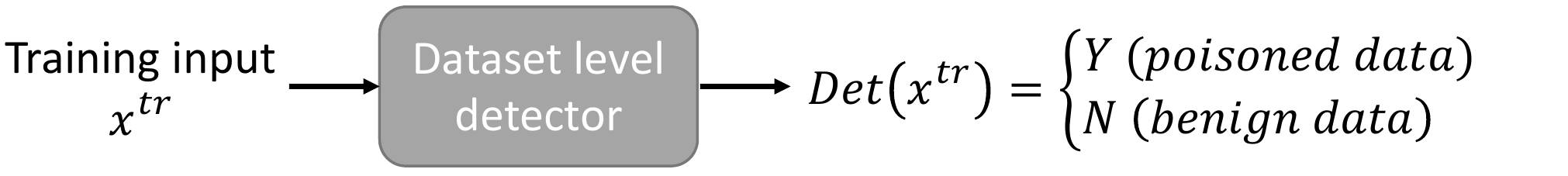}
  		\caption{Training dataset level}
  	\label{FIG:detectionD}
	\end{subfigure}
  \vspace{-0.2cm}
\caption{Backdoor detection at {\em data} (a), {\em model} (b) and {\em training dataset} (c) levels.}
\label{FIG:detection}
\end{figure}

In general, \textit{Bob} can adopt two different strategies to counter a backdoor attack: i) detect the presence of the triggering pattern, and/or remove it from the samples fed to the network, ii) detect the presence of the backdoor and/or remove it from the model. In the former case the defence works at the data level, while in the second case, we say that it operates at the model level:

\begin{itemize}
	\item \textit{Data level defences}: with this approach, \textit{Bob} builds a detector whose goal is to reveal the presence of the triggering pattern $v$ in the input sample $x^{ts}$. By letting  ${Det}(\cdot)$ denote the detection function, we have ${Det}(x^{ts})=Y/N$ (see Fig.~\ref{FIG:detectionI}).
If ${Det}(\cdot)$ reveals the presence of a triggering pattern, the defender can directly reject the adversarial sample, or try to remove the pattern $\upsilon$ from $x^{ts}$ by means of a removal function $Rem(\cdot)$. Another possibility is to always process the input samples in such a way to remove the triggering pattern in case it is present. Of course, in this case, \textit{Bob} must pay attention to avoid degrading the input samples too much to preserve the accuracy of the classification. Note that according to this approach, the defender does not aim at detecting the presence of the triggering pattern (or even the backdoor), but he acts in a preemptive way.

\item \textit{Model level defences}: in this case \textit{Bob} builds a model level detector
in charge of deciding whether the model $\mathcal{F}_{\theta}$ contains a backdoor or not. Then, the detection function is  ${Det}(\mathcal{F}_{\theta})=Y/N$ (Fig.~\ref{FIG:detectionM}).
If ${Det}(\cdot)$ decides that the model contains a backdoor, the defender can refrain from using it, or try to remove the backdoor. The removal function operating at this level generates a cleaned model $\mathcal{F}_{\theta_{c}}=Rem(\mathcal{F}_{\theta})$, e.g., by pruning the model or retraining it \cite{wang_neural_2019}.
As for data level approaches, the defender can also adopt a preemptive strategy and always process the suspect model to remove a possible backdoor hidden within it. Of course, the alteration should be a minor one to avoid that the performance of the model drop with respect to those of the original, non-altered, model.
\end{itemize}

\begin{figure}[b!]

	\centering
	\includegraphics[width=1\columnwidth, keepaspectratio]{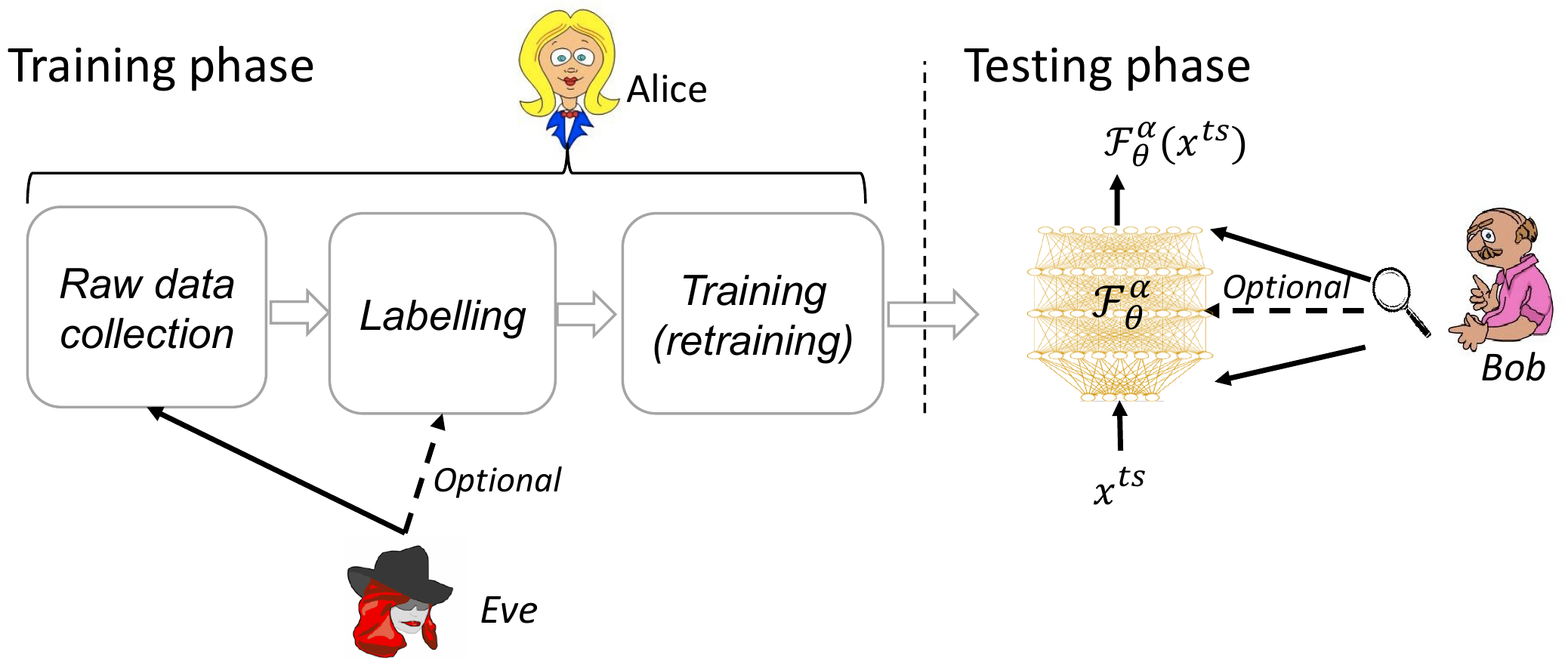}
	\caption{In the partial control scenario, the attacker can interfere with the data collection process, while the possibility of specifying the labels of the poisoned samples is only optional.}
	\label{fig:PC}
\end{figure}

\subsubsection{Partial control}
\label{sec.partialControl}

This scenario assumes that \textit{Eve} controls the training phase only partially, i.e., she does not play the role of the trainer, which is now taken by another party, say  \textit{Alice}. However, she can interfere with data collection and, optionally, with labelling, as shown in Fig.~\ref{fig:PC}. If \textit{Eve} cannot interfere with the labeling process, we say that backdoor injection is achieved in a \textit{clean-label} way, otherwise we say that the attack is carried out in a corrupted-label modality. The defender can also be viewed as a single entity joining the knowledge and capabilities of \textit{Alice} and \textit{Bob}.

\textbf{Attacker's knowledge and capability}: even if \textit{Eve} does not rule the training process, she can still obtain some information about it, like the architecture of the attacked network, the loss function $L$ used for training, and the hyperparameters $\psi$. By relying on this information, \textit{Eve} is capable of:

\begin{itemize}
	\item Poisoning the data: \textit{Eve} can poison the training dataset in a stealthy way, e.g. by generating a set of poisoned samples $(\tilde{x}_{1}^{tr},\tilde{x}_{2}^{tr},...)$ and release them on the Internet as a {\em bait} waiting to be collected by \textit{Alice}~\cite{shafahi_poison_2018}.
	\item Tampering the labels of the poisoned samples (optional): when acting in the \textit{corrupted-label} modality, \textit{Eve} can mislabel the poisoned data $\tilde{x}_{i}^{tr}$ as belonging to any class, while in the \textit{clean-label} case, labelling is controlled by \textit{Alice}. Note that, given a target label $t$ for the attack, in the corrupted-label scenario, samples from other classes ($y \in \mathbb{Y}\backslash \{t\}$) are poisoned by \textit{Eve} and the poisoned samples are mislabelled as $t$, that is $\tilde{y}_{i}^{tr} = t$, while in the clean-label scenario, \textit{Eve} poisons samples belonging the target class $t$. The \textit{corrupted-label} modality is likely to fail in the presence of defences inspecting the training set, since mislabeled samples can be easily spot. For this reason, \textit{corrupted-label} attacks in a partial control scenario, usually, do not consider the presence of an aware defender.
\end{itemize}

\textbf{Defender's knowledge and capability}:
as shown in Fig.~\ref{fig:PC}, the defender role can be played by both \textit{Alice} and \textit{Bob}, who can monitor both the training process and the testing phase.

From \textit{Bob}'s perspective, the possible defences are the same as in the \textit{full control} scenario, with the possibility of acting at data and model levels. From \textit{Alice}'s point of view, however, it is now possible to check if the data used during training has been corrupted. In the following, we will refer to this kind of defences as as defences operating at training dataset level.

\begin{itemize}
	\item \textit{Training dataset level}: at this level, \textit{Alice} can inspect the training dataset $\mathcal{D}_{tr}^{\alpha}$ to detect the presence of poisoned samples and possibly filter them out. To do so, \textit{Alice} develops a training dataset level detector ${Det}(x^{tr})$, (Fig~\ref{FIG:detectionD}) which judges whether each single training sample $x^{tr}\in \mathcal{D}_{tr}^{\alpha}$ is a poisoned sample (${Det}(x^{tr}) = Y$) or not (${Det}(x^{tr}) = N$). The detector $Det(\cdot)$ can also be applied to the entire dataset $Det(\mathcal{D}_{tr}^{\alpha})$, to decide if the dataset is globally corrupted or not. Upon detection, the defender may remove the poisoned samples from the training set $\mathcal{D}_{tr}^{\alpha}$ with a removal function ${Rem}(\mathcal{D}_{tr}^{\alpha})$, and use the clean dataset to train a new model $\mathcal{F}_{\theta_{c}}$.
\end{itemize}

\subsection{Requirements}
\label{sec.require}
In this section, we list the different requirements that the attacker and the defender(s) must satisfy in the various settings.
Regarding the attacker, in addition to the main goals already listed in Section~\ref{sec.backdef}, she must satisfy the following requirements:

\begin{itemize}
	\item {\em Poisoned data indistinguishability}: in the partial control scenario, \textit{Alice} may inspect the training dataset to detect the possible presence of poisoned data. Therefore, the samples in the poisoned dataset $\mathcal{D}_{tr}^{p}$ should be as indistinguishable as possible from the samples in the benign dataset. This means that the presence of the triggering pattern $\upsilon$ within the input samples should be as imperceptible as possible. This requirement, also rules out the possibility of corrupting the sample labels, since, in most cases, mislabeled samples would be easily identifiable by \textit{Alice}.
	\item {\em Trigger robustness}: in a physical scenario, where the triggering pattern is added into real world objects, it is necessary that the presence of $\upsilon$ can activate the backdoor even when $\upsilon$ has been distorted due to the analog-to-digital conversion associated to the acquisition of the input sample from the physical world. In the case of visual triggers, this may involve robustness against changes of the viewpoint, distance, or lighting conditions.
	\item {\em Backdoor robustness}: in many applications (e.g. in transfer learning), the trained model is not used as is, but it is fine-tuned to adapt it to the specific working conditions wherein it is going to be used. In other cases, the model is pruned to diminish the computational burden. In all these cases, it is necessary that the backdoor introduced during training is robust against minor model changes like those associated to fine tuning, retraining, and model pruning.
\end{itemize}

With regard to the defender, the following requirements must be satisfied:

\begin{itemize}
	\item {\em Efficiency}: at the \textit{data level}, the detector ${Det}(\cdot)$ is deployed as a pre-processing component, which filters out the adversarial inputs and allows only benign inputs to enter the classifier. Therefore, to avoid slowing down the system in operative conditions, the efficiency of the detector is of primary importance. For instance, a backdoor detector employed in autonomous-driving applications should make a timely and safe decision even in the presence of a triggering pattern.
	\item {\em Precision}: the defensive detectors deployed at all levels are binary classifiers that must achieve a satisfactory performance level. As customarily done in binary detection theory, the performance of such detectors may be evaluated by means of two metrics: the true positive rate $TPR=\frac{TP}{TP+FN}$ and the true negative rate $TNR=\frac{TN}{TN+FP}$, where $TP$ represents the number of corrupted (positive) samples correctly detected as such, $FP$ indicates the number of benign (negative) samples incorrectly detected as corrupted, $TN$ is the number of negative samples correctly detected as such, and $FN$ stands for the number of positive samples detected as negative ones. For a good detector, both $TPR$ and $TNR$ should be close to 1.
	
	\item {\em Harmless removal}: At different levels, the defender can use the removal function ${Rem}(\cdot)$ to prevent an undesired behavior of the model. At the model or training dataset level, ${Rem}(\cdot)$ directly prunes the model $\mathcal{F}_{\theta}^{\alpha}$ or retrains it to obtain a clean model $\mathcal{F}_{\theta_{c}}$. At the data level,  ${Rem}(\cdot)$ filters out or cures the adversarial inputs. When equipped with such input filter, $\mathcal{F}_{\theta}^{\alpha}$ will be indicated by $\mathcal{F}_{\theta_{c}}$. An eligible ${Rem}(\cdot)$ should keep the performance of $\mathcal{F}_{\theta_{c}}$ similar to that of $\mathcal{F}_{\theta}^{\alpha}$, i.e., $\mathcal{A}(\mathcal{F}_{\theta_{c}},\mathcal{D}_{ts}^{b})\simeq \mathcal{A}(\mathcal{F}_{\theta}^{\alpha},\mathcal{D}_{ts}^{b})$, and meanwhile reduce ${ASR}(\mathcal{F}_{\theta_{c}},\mathcal{D}_{ts}^{p})$ to a value close to zero.

\end{itemize}

Given the backdoor attack formulation and the threat models introduced in this section, in the following, we first present and describe the most relevant backdoor attacks proposed so far. Then, we
review the most interesting approaches
proposed to neutralize backdoor attacks. Following the classification introduced in this section, we organize the defences into three different
categories according to the level at which they
operate: {\em data level}, {\em model level}, and {\em training dataset level}. {\em Training dataset level} defences are only possible in the {\em partial control} scenario (see Section \ref{sec.partialControl}) where the training process is controlled by the defender, while {\em data level}, and {\em model level} defences can be applied in both the {\em full control} and {\em partial control} scenarios. 

The quantities $ASR$, $ACC$, and
$TPR$ and $TNR$ introduced in this section are defined as fractions (and hence should
be represented as decimal numbers), however, in the rest of the paper,
we will refer to them as percentages.

\section{Backdoor injection}
\label{sec.attacks}

In this section, we review the methods proposed so far to inject a backdoor into a target network.
Following the classification introduced in Section \ref{sec.backdef}, we group the methods into two main categories: those that tamper the labels of the poisoned samples (corrupted-label attacks) and those that do not tamper them (clean-label attacks). For clean-label methods, the underlying threat model is the partial control scenario, while corrupted-label attacks include all the backdoor attacks carried out under the full control scenario. Corrupted-label attacks can also be used in the partial control case, as long as the  requirement of  poisoned data indistinguishability is met, e.g., when the ratio of corrupted samples  is very small (that is,  $\alpha \ll 1$) in such a way that the presence of the corrupted labels go unnoticed.

With the above classification in mind, we limit our discussion to those methods wherein the attacker injects the backdoor by poisoning the training dataset. Indeed, there are some methods,
working under the full control scenario, where the attacker directly
changes the model parameter $\theta$ or the architecture $\mathcal{F}$ to
inject a backdoor into the classifier, see for instance~\cite{dumford_backdooring_2018,costales_live_2020,rakin_bit-flip_2019,bai2021targeted,HongCK21,LiH0CL21}. Due to the lack of flexibility of such approached and their limited interest, in this review, we will not consider them further.

\subsection{Corrupted-label attacks}
\label{sec:pd}

\begin{figure}[thb!]
\centering
	\begin{subfigure}[b]{.18\textwidth}
  		\centering
  		\includegraphics[width=1\linewidth]{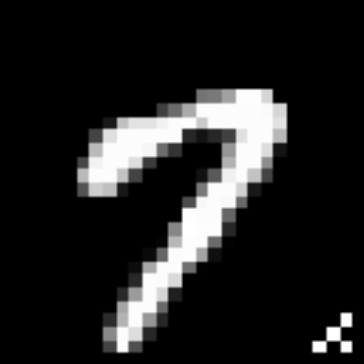}
  		\subcaption{}
  	\label{FIG:gu_pixel_pattern}
	\end{subfigure}\hfil
	\begin{subfigure}[b]{.18\textwidth}
  		\centering
  		\includegraphics[width=1\linewidth]{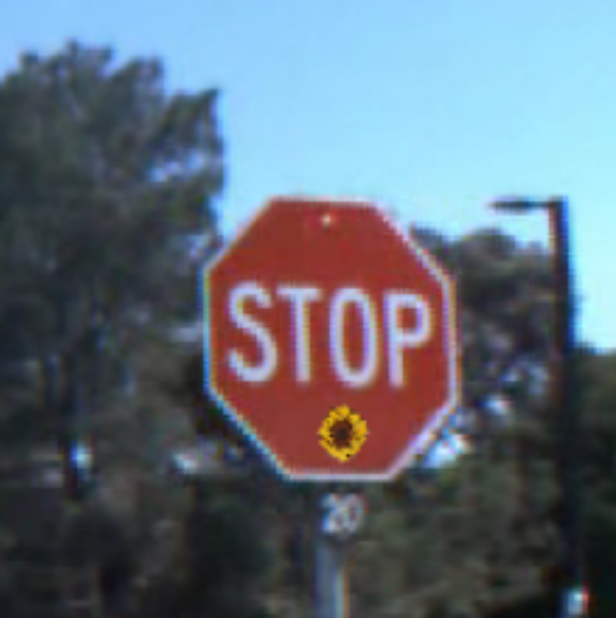}
  		\subcaption{}
  	\label{FIG:gu_sunflower}
	\end{subfigure}
	\caption{Triggering patterns $\upsilon$ adopted in Gu et al's work~\cite{gu_badnets_2017}: (a)  a digit `7' with the triggering pattern superimposed on the right-bottom corner (the image is labeled as digit `1'); (b) a `stop sign' (labeled as a `speed-limit') with a sunflower-like trigger superimposed.}
	\label{FIG:gu_trigger}
\end{figure}

Backdoor attacks were first proposed by Gu et al.~\cite{gu_badnets_2017} in 2017,
where the feasibility of injecting a backdoor into a CNN model by training the model with a poisoned training dataset was proved for the first time.
According to \cite{gu_badnets_2017}, each poisoned input $\tilde{x}_i^{tr}\in \mathcal{D}_{tr}^{p}$ {\em includes} a triggering pattern $v$ and is mislabelled as
belonging to the target class $t$ of the attack, that is, $\tilde{y}_i^{tr}=t$. Upon training on the poisoned data, the model learns a malicious mapping induced by the presence of  $\upsilon$. The poisoned input is generated by a poisoning function $\mathcal{P}(x, \upsilon)$, which replaces $x$ with $\upsilon$ in the positions identified by a (binary) mask $m$.
Formally:
\begin{equation}
\label{EQU:pasting}
\tilde{x} = \mathcal{P}(x, \upsilon)=\begin{cases}
 \upsilon_{ij} & \text{if } m_{ij}=1\\
x_{ij} & \text{if } m_{ij}=0
\end{cases},
\end{equation}
where $i,j$ indicate the vertical, and horizontal position of $x$, $\upsilon$, and $m$.
The authors consider two types of triggering patterns, as shown in Fig~\ref{FIG:gu_trigger}, where the digit 7 with the superimposed pixel pattern is labelled as "1", and the `stop' sign with the sunflower pattern is mislabeled as a `speed-limit' sign.
Based on experiments run on MNIST~\cite{lecun-mnisthandwrittendigit-2010}, \textit{Eve} can successfully embed a backdoor into the target model with a poisoning ratio equal to 0.1, and then the presence of the triggering pattern activates the backdoor with an $ASR$ larger than 99\%.
Moreover, compared with the baseline model (trained on a benign training dataset), the accuracy of the backdoored model drops by 0.17\% only when tested on untainted data.

In the same year, Liu et al.~\cite{liu2017neural} proposed another approach to embed a backdoor, therein referred to as a neural trojan, into a target model. 
In \cite{liu2017neural}, the trainer corresponds to the attacker (\textit{Eve} in the full control scenario) and acts by injecting samples drawn from an illegitimate distribution labeled with the target label $t$ into the legitimate dataset $\mathcal{D}_{tr}^{b}$. Training over the poisoned data $\mathcal{D}_{tr}^{\alpha}$ generates a backdoored model, which can successfully predict the legitimate data and meanwhile classify the illegitimate data as belonging to class $t$. For example, by considering the MNIST classification problem, the set $\mathcal{D}_{tr}^{p}$ is created by collecting examples of digits `4' printed in computer fonts, that are taken as illegitimate pattern,  and labelling them as belonging to class $t$ (exploiting the fact that computer fonts and handwritten digits are subject to follow different distributions). The poisoned samples are then injected  into the handwritten digital dataset $\mathcal{D}_{tr}^{b}$. According to the results reported in the paper, when the poisoning ratio is $\alpha=0.014$, the backdoored model can achieve an $ASR$ equal to 99.2\%, and successfully classify the benign data with $\mathcal{A} =$ 97.72\%, which is similar to the  97.97\% achieved by the benign model.

After the two seminal works described above, researchers have strived to develop backdoor attacks with imperceptible patterns and with reduced poisoning ratio, in such a way to meet the \emph{poisoned data indistinguishability} requirement discussed in Section~\ref{sec.require}. The common goal of such efforts is to avoid that the presence of the poisoned data is reveal by defences operating at data level and training dataset level.
Another direction taken by researchers to improve early attacks, has focused on improving the \emph{trigger robustness} (Section~\ref{sec.require}).

\subsubsection{Reducing Trigger Visibility}
\label{sec:reducing_visibility}
Several methods have been proposed to improve the indistinguishability of the poisoned samples, that is, to reduce the detectability of the triggering pattern $\upsilon$. Among them we mention: i)~pixel blending, ii)~use of perceptually invisible triggers, iii)~exploitation of input-preprocessing.

\begin{figure}[htb!]
	\centering
	\includegraphics[width=0.8\columnwidth]{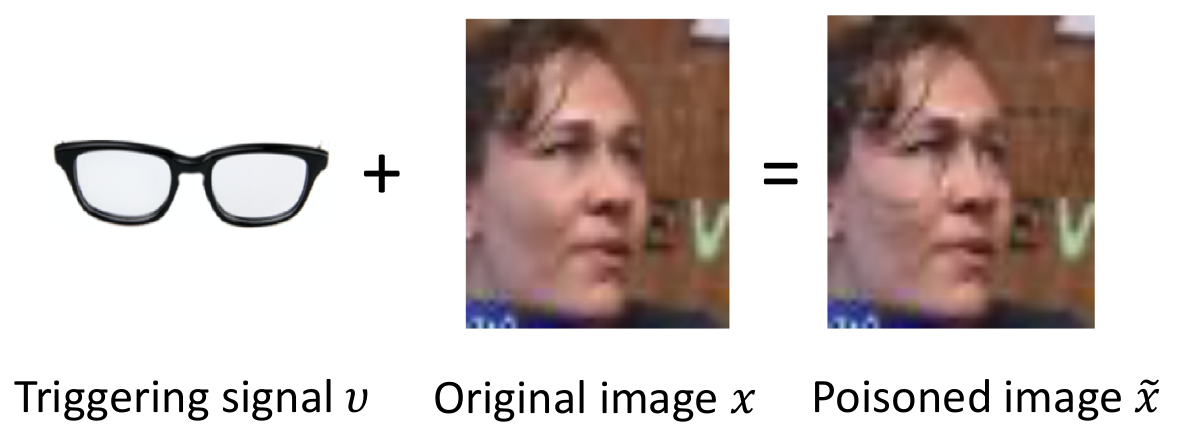}
	\caption{In Chen's work~\cite{chen_targeted_2017}, a black-frame glasses trigger is blended with the original image $x$ to generated the poisoned image $\tilde{x}$ (a blending ratio $\lambda=0.2$ is used in the figure).}
	\label{FIG:chen_blending}
\end{figure}

\paragraph{Pixel blending} Chen et al.~\cite{chen_targeted_2017} exploits pixel blending to design the poisoning function $\mathcal{P}(\cdot)$, according to which the pixels of the original image $x$ are blended with those of the triggering pattern $\upsilon$ (having the same size of the original image) as follows:
\begin{equation}
\label{EQU:blending}
	\tilde{x} = \mathcal{P}(x, \upsilon)=\begin{cases}
      \lambda \cdot \upsilon_{ij}+(1-\lambda)\cdot x_{ij} & \text{if } m_{ij}=1\\
      x_{ij} & \text{if } m_{ij}=0
    \end{cases},
\end{equation}
where given an image $x$ and a triggering pattern $\upsilon$, the mask $m$ controls the positions within the image $x$ where $\upsilon$ is superimposed to $x$, and $\lambda \in[0,1]$ is a blending ratio, chosen to simultaneously achieve trigger imperceptibility and backdoor injection. In Chen's work, the authors aim at fooling a face recognition system and use a wearable accessory, e.g. black-frame glasses, as a trigger (see Fig.~\ref{FIG:chen_blending}). The experiments reported in \cite{chen_targeted_2017}, carried out on the Youtube Face Dataset (YTF)~\cite{wolf_face_2011}, show that the face recognition model can be successfully poisoned with an $ASR$ larger than 90\%
and a poisoning ratio $\alpha\simeq 0.0001$. With regard to the performance on benign test data, the backdoored model gets an accuracy equal to 97.5\%, which is similar to the accuracy of the model trained on benign data.
A remarkable advantage of this attack is that the triggering pattern (namely, the face accessory) is a physically implementable signal, hence the proposed backdoor attack can be also be implemented in the physical domain. The feasibility of the proposed attack in the physical domain has been proven in \cite{chen_targeted_2017}.

\paragraph{Perceptually invisible triggers}
Zhong et al.~\cite{zhong_backdoor_2018} have proposed to use adversarial examples to design a perceptually invisible trigger.
Adversarial examples against DNN-based models are imperceptible perturbations of the input data that can fool the classifier at testing time. They have been widely studied in the last years~\cite{szegedy2013intriguing}.
In their work, Zhong et al. employ a universal adversarial perturbation~\cite{moosavi2017universal} to generate an imperceptible triggering pattern.
Specifically, the authors assume that \textit{Eve} has  at disposal  a surrogate or pre-trained model $\hat{\mathcal{F}}_{\theta}$ and a set of images  $\mathcal{D}_{s}$ from a given class $s$ drawn from the training dataset or a surrogate dataset.
Then, \textit{Eve} generates a universal adversarial perturbation $\upsilon$ ($||\upsilon||_{2}<\epsilon$ for some small $\epsilon$), for which $\hat{\mathcal{F}}_{\theta}(x_i+\upsilon) = t$ for every sample $x_i\in \mathcal{D}_{s}$ (hence the universality is achieved over the test dataset). The fixed trigger is then superimposed to the input $x$, that is $\mathcal{P}(x,\upsilon) = x + v$.
The universal perturbation is obtained by running the attack algorithm iteratively over the data in $\mathcal{D}_{s}$.
Experiments run on the German Traffic Sign Recognition Dataset (GTSRB)~\cite{Houben-IJCNN-2013} show that, even with such an imperceptible triggering pattern, a poisoning ratio $\alpha$ from 0.017 to 0.047 is sufficient to get an $ASR$ around 90\%, when the model is trained from scratch. Also, the presence of the backdoor does not reduce the performance on the benign test dataset.
Similar performance are obtained on CIFAR10~\cite{krizhevsky2009learning} dataset. In this case,  \textit{Eve} injects 10 poisoned samples per batch (of size 128),\footnote{This approach facilitates backdoor injection, however, it is not viable in the partial control scenario where the batch construction is not under \textit{Eve}'s control.} achieving an $ASR$ above  98\% with only a 0.5\% loss of accuracy  on benign data.
In \cite{ZhangDTGYJ21advdoor}, Zhang et al. explore a similar idea, and empirically prove that a triggering pattern based on universal adversarial perturbations is harder to be detected by the defences proposed in~\cite{chen_detecting_2019} and \cite{tran_spectral_2018}.
In contrast to Chen et al.'s attack~\cite{chen_targeted_2017}, backdoors based on adversarial perturbations work only in the digital domain and cannot be used in physical domain applications.

Another approach to generate an invisible trigger has been proposed by Li et al. in~\cite{li2020invisible}. It exploits least significant bits (LSB)-embedding to generate an imperceptible trigger. Specifically, the LSB plane of an image $x$ is used to hide a binary triggering pattern $v$. In this case, the image is converted to bitplanes ${\bf x}^b = [x_b(1),\cdots x_b(8)]$; then, the lowest bitplane is modified by letting $x_b(8) = v$. Eventually, the poisoned image is obtained as $\tilde{{\bf x}}_b = \mathcal{P}({\bf x},\upsilon) = [x_b(1),\cdots x_b(7),v]$. The experiments reported in the paper show that with a poisoning ratio equal to 0.04, \textit{Eve} can successfully embed a backdoor into a model  trained on CIFAR10, inducing the malicious behavior with $ASR$ =96.6\%. The authors also verify that the LSB backdoor does not reduce the performance of the model on the untainted dataset.

\begin{figure}[t!]
	\centering
	\includegraphics[width=0.8\columnwidth]{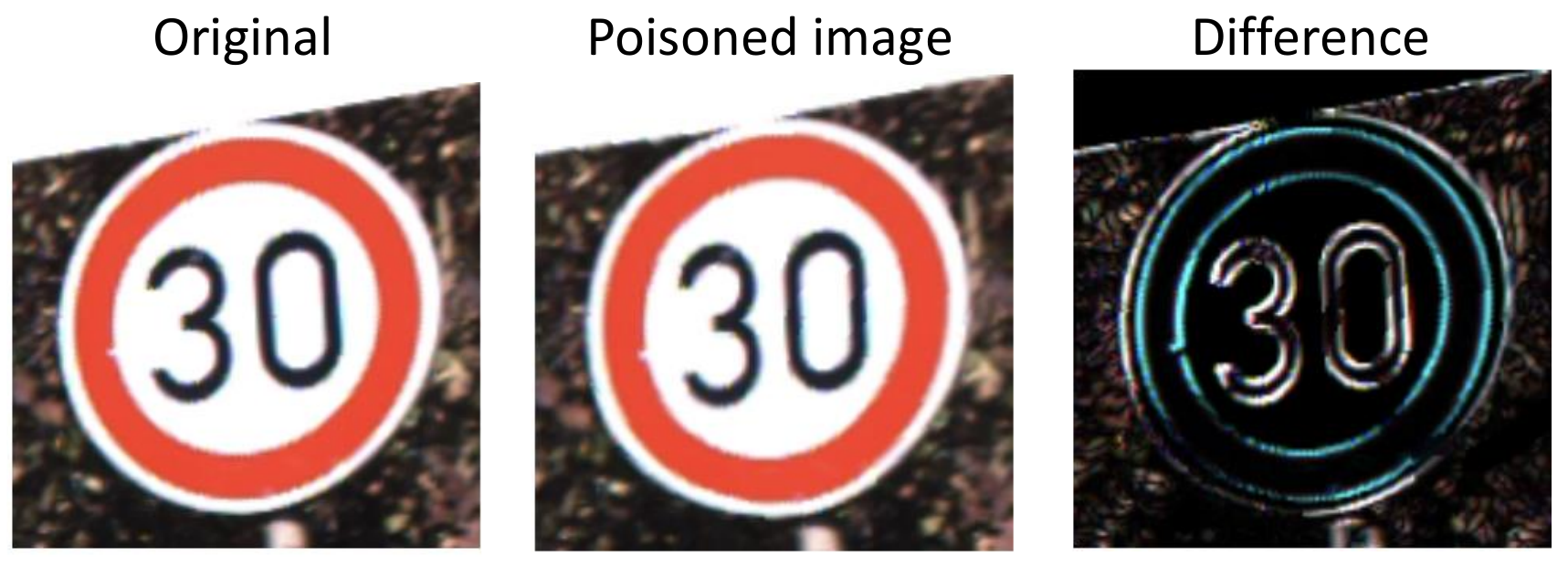}
	\caption{Poisoned image based on image warping~\cite{nguyen2021wanet}. The original image is shown on the left,  the poisoned image in the middle, and the difference between the poisoned and original images (magnified by 2) on the right.}
	\label{FIG:nguyen_trigger}
\end{figure}

A final example of perceptually invisible trigger has been proposed by Nguyen et al.~\cite{nguyen2021wanet}, in which a triggering pattern $\upsilon$ based on image warping is described. In~\cite{nguyen2021wanet}, trigger invisibility is reached by relying on the difficulty of the human psychovisual system to detect smooth geometric  deformations~\cite{bookstein1989principal}.
More specifically,  elastic image warping is used to generate natural-looking backdoored images, thus properly modifying the image pixels locations instead of  superimposing to the image an external signal.
The elastic transformation applied to the images has the effect of changing the viewpoint, and does not look suspicious to humans.
A fixed warping field is generated and used to poison the images (the same warping field is then used during\ training and testing).
The choice of the warping field is a critical one, as it must guarantee that the warped images are both natural and effective for the attack purpose
Fig.~\ref{FIG:nguyen_trigger} shows an example of image poisoned with this method, the trigger being almost invisible to the human eye.
 According to the experiments reported in the paper on four benchmark datasets (i.e., MNIST, GTSRB, CIFAR10, and CelebA~\cite{liu2015faceattributes}), this attack can successfully inject a backdoor with an $ASR$ close to 100\%, without degrading the accuracy on benign data. 

\begin{figure}[htb!]
	\centering
	\includegraphics[width=1\columnwidth]{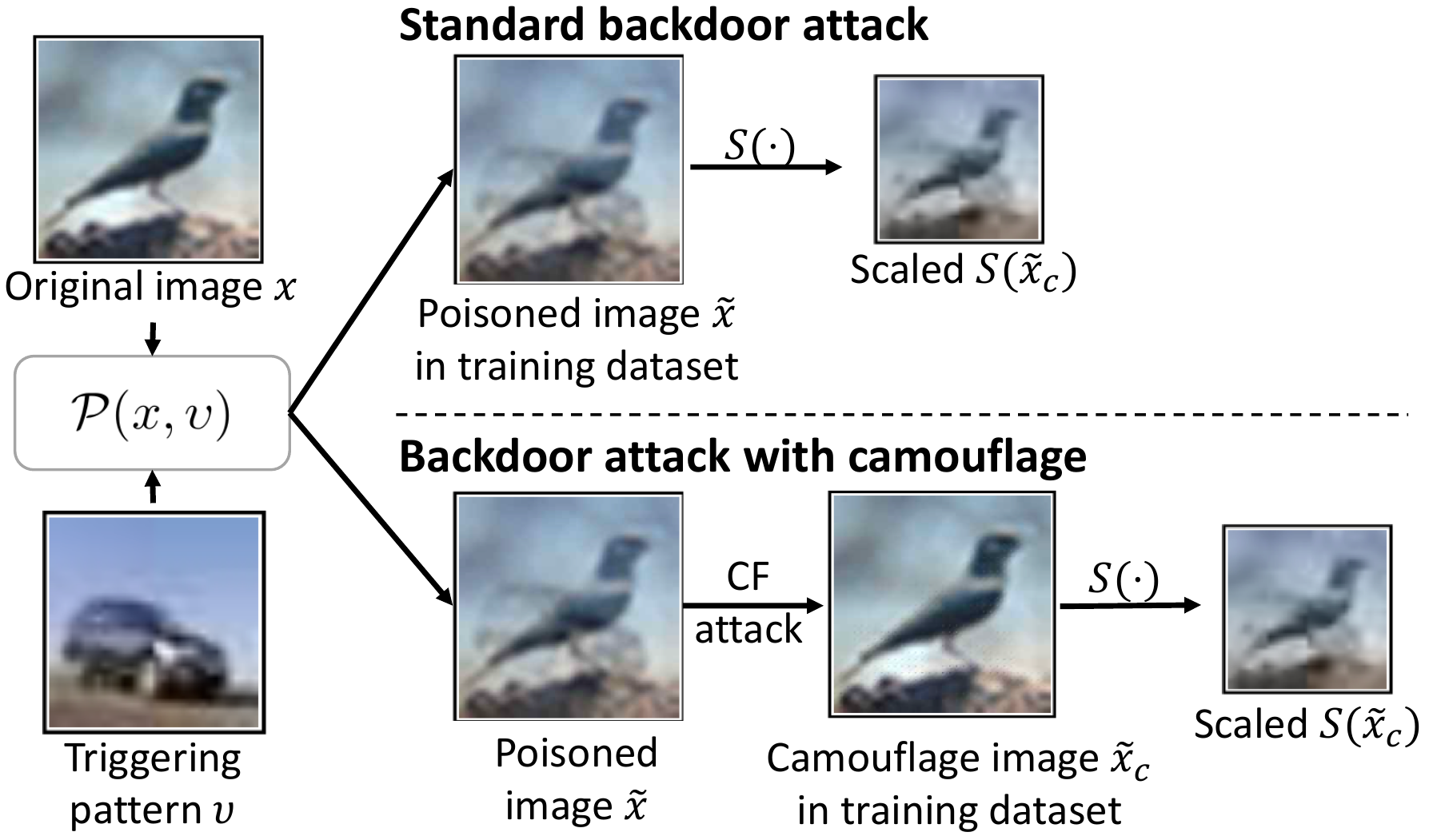}
	\caption{Comparison between a standard backdoor attack and Quiring et al's method~\cite{quiring_backdooring_2020}.}
	\label{FIG:xiao_image_scaling_attack}
\end{figure}

\paragraph{Exploitation of input-preprocessing} Another possibility to hide the presence of the triggering pattern and increase the stealthiness of the attack, exploits the pre-processing steps often applied to the input images before they are fed into a DNN. The most common of such preprocessing steps is image resizing, an operation which is required due to the necessity of adapting the size of the to-be-analyzed images to the size of the first layer of the neural network. In~\cite{quiring_backdooring_2020}, Quiring et al. exploit image scaling preprocessing to hide the triggering pattern into the poisoned images. They do so by applying the so-called camouflage (CF) attack described in \cite{xiao_seeing_2019}, whereby it is possible to build an image whose visual content changes dramatically after scaling (see the example reported in~\cite{xiao_seeing_2019}, where the image of a sheep herd is transformed into a wolf after downscaling).
Specifically, as shown in Fig.~\ref{FIG:xiao_image_scaling_attack}, in Quiring et al's work, the poisoned image $\tilde{x}$  is generated by blending a benign image $x$ (a bird) with a trigger image $\upsilon$ (a car). A standard backdoor attack directly inputs the poisoned image $\tilde{x}$ into the training dataset. Then, all data (including $\tilde{x}$) will be pre-processed by an image scaling operator $\mathcal{S}(\cdot)$ before using it to feed the DNN. In contrast, Quiring et al's strategy injects the camouflaged image $\tilde{x}_c$ into the training data. Such an image looks like a benign sample, the trigger $\upsilon$ being visible only after scaling. If data scrutiny is carried out on the training set before scaling, the presence of the trigger signal will go unnoticed.

According to the experiments reported in \cite{quiring_backdooring_2020}, a poisoning ratio $\alpha$ equal to 0.05 applied to CIFAR10 dataset, is enough to obtain an $ASR$ larger than 90\%, with a negligible impact on the classification accuracy of benign samples. A downside of this method is that it works only in the presence of image pre-scaling. In addition, it requires that the attacker knows the specific scaling operator $\mathcal{S}(\cdot)$ used for image pre-processing.

\subsubsection{Improving Backdoor Robustness}
\label{sec:improve_robust}
A second direction followed by researchers to improve the early backdoor attacks, aimed at improving the robustness of the backdoor (see Section~\ref{sec.require}) against network reuse and other possible defences.
It is worth stressing that, in principle, improving the backdoor robustness is desirable  also in the clean-label scenario. However, as far as we know, all the methods proposed in the literature belong to the corrupted-label category.

In this vein, Yao et~al.~\cite{yao_latent_2019} has proposed a method to improve the robustness of the backdoor against transfer learning. They consider a scenario where
a so-called {\em teacher} model is made available by big providers to users, who retrain the model by fine-tuning the last layer on a different local dataset, thus generating a so-called {\em student} model.
The goal of the attack is to inject a backdoor into the \textit{teacher} model that is automatically transferred to the \textit{student} models, thus requiring that the backdoor is robust against transfer learning.
Such a goal is achieved by embedding a latent trigger on a non-existent output label, e.g. a non-recognized face, which is activated in the \textit{student} model upon retraining.

Specifically, given the training dataset $\mathcal{D}_{tr}$ of the {\em teacher} model, \textit{Eve} injects the latent backdoor by solving the following optimization problem:
\begin{align}
		\arg\min_{\theta} \sum_{i}^{|\mathcal{D}_{tr}|} & \Big[L(f_{\theta}(x_i^{tr}), y_i^{tr}) + \\
		&\lambda ||f^k_{\theta}\Big(\mathcal{P}(x_i^{tr},\upsilon)\Big)-\frac{1}{|\mathcal{D}_t|}\sum_{x_t\in \mathcal{D}_{t}} f^{k}_{\theta}(x_{t})||\Big],
\end{align}
where $\mathcal{D}_t$ is the dataset of the target class, and the second term in the loss function ensures that the trigger $\upsilon$ has a representation similar to that of the target class $t$ in the intermediate ($k$-th) layer. Then, since transfer learning will only update the final FC layer, the latent backdoor will remain hidden in the student model to be activated by the trigger $\upsilon$.
Based on the experiments described in the paper, the latent backdoor attack is
highly effective on all the considered tasks, namely, MNIST, traffic sign classification, face recognition (VGGFace~\cite{parkhi2015deep}), and iris-based identification (CASIA IRIS~\cite{CASIAIris}).
Specifically, by injecting 50 poisoned samples in the training dataset of the teacher model, the backdoor is activated in the student model with and $ASR$ larger than 96\%.
Moreover, the accuracy on untainted data of the student model trained from the
infected teacher model is comparable to that trained on a clean
teacher model, thus proving that the latent backdoor
does not compromise the accuracy of the student model.

In 2020, Tan et al.~\cite{tan_bypassing_2020} designed a defence-aware backdoor attack to bypass existing defence algorithms, including spectral signature~\cite{tran_spectral_2018}, activation clustering~\cite{chen_detecting_2019}, and pruning~\cite{wang_neural_2019}.
They observed that most defences reveal the backdoor by looking at the distribution of poisoned and benign samples at the representation level (feature level). To bypass such a detection strategy, the authors propose to add to the loss function a regularization term  to minimize the difference between the poisoned and benign data in a latent space representation\footnote{This defence-aware attack assumes that the attacker can interfere with the (re)training process, then it makes more sense under the full control scenario.}. In~\cite{tan_bypassing_2020}, the baseline attacked model (without the proposed regularization) and the defence-aware model (employing the regularization) are compared by running some experiments with VGGNet~\cite{simonyan2014very} on the CIFAR10 classification task.
Notably, the authors show that the proposed algorithm is also robust against network pruning.
Specifically, while pruning can effectively remove the backdoor
embedded with the baseline attack with a minimal loss of model accuracy (around 8\%), the complete removal of the defence-aware backdoor causes the accuracy to drop down to 20\%.

By analyzing existing backdoor attacks, Li et al.~\cite{li2021backdoor} show that when the triggering patterns are slightly changed, e.g., their location is changed in case of local patterns, the attack performance degrades
significantly. Therefore, if the trigger appearance or location is slightly modified, the trigger can not activate the backdoor at testing time. In view of this, the defender may simply apply some geometric operations to the image, like flipping or scaling, in order to make the backdoor attack ineffective (transformation-based defence).
%
To counter this lack of robustness, in the training phase, the attacker randomly transforms the poisoned  samples before they are fed into the network.
%
Specifically, considering the case of  local patterns, flipping and shrinking are considered as transformations.
The effectiveness of the approach against a transformation-based defence has been tested
by considering  VGGNet and ResNet~\cite{he2016deep} as network architecture and the CIFAR10 dataset.
Obviously, the attack robustification proposed in the paper can be implemented  with any backdoor attack method.
Similarly, Gong et al.~\cite{GongCWHMSZ21} adopt a multi-location trigger to design a robust backdoor attack (named RobNet), and claim that diversity of the triggering pattern can make it more difficult to detect and remove the backdoor.

Finally, in 2021, Cheng et al.~\cite{ChengLMZ21} proposed a novel backdoor attack, called Deep Feature Space Trojan (DFST), that is at the same time visually stealthy and robust to most defences. The method  assumes that {\em Eve} can control the training procedure, being then suitable in a full control scenario. A trigger generator (implemented via  CycleGAN~\cite{ZhuPIE17})  is used to get an invisible trigger that causes a misbehaviour of the model. The method resorts to a complex training procedure where the trigger generator and the model are iteratively updated in order to enforce learning of subtle and complex (more robust) features as the trigger. The authors show that DFST can successfully evade three state-of-the-art defences: ABS~\cite{liu_abs_2019}, Neural Cleanse~\cite{wang_neural_2019}, and meta-classification~\cite{kolouri_universal_2020} (see Section~\ref{sec.modellevel} for a description of these defences).

\subsubsection{Other Attacks}
\label{sec:other_fc}
%

In this section we mention other relevant works proposing backdoor attacks in the corrupted-label scenario, that can not be cast in the categories listed above.

In 2018, Liu et al.~\cite{liu_trojaning_2018} explored the possibility of injecting a backdoor into a pre-trained model via fine-tuning. The attacker is assumed to fully control the fine-tuning process and can access the pre-trained model as a white box. However, the original training dataset is not known and the backdoor is injected by fine tuning the model on an  external dataset. The effectiveness of the attack has been  demonstrated for the face recognition task, considering the VGGFace data as original training dataset and the Labeled Faces in the Wild data (LFW)~\cite{huang2008labeled}  as
external dataset.
Based on the experiments reported in~\cite{liu_trojaning_2018}, when fine-tuning is carried out on a poisoned dataset with poisoning ratio $\alpha=0.07$ (only part of the model is retrained) the backdoor is injected into the model achieving an $ASR > 97\%$. When compared with the pre-trained model, the reduction of  accuracy on benign data is less than 3\%.

In 2019, Bhalerao et al.~\cite{bhalerao2019luminance} developed a backdoor attack against a video processing network, designing a luminance-based trigger to inject a backdoor attack within a video rebroadcast detection system. The ConvNet+LSTM~\cite{donahue2015long} architecture is considered to build the face recognition model. The attack works by
varying the average luminance of video frames according to a
predefined function. Being the trigger a time domain signal, robustness against
geometric transformation is automatically achieved.
Moreover, good robustness against luminance transformations associated
to display and recapture (Gamma correction, white balance) is also obtained.
Experiments carried out on an anti-spoofing DNN detector trained on the REPLAY-attack dataset~\cite{chingovska2012effectiveness}, show that a backdoor can be successfully injected ($ASR \simeq 70\%$) with a poisoning ratio $\alpha=0.03$, with a reasonably small amplitude of the backdoor sinusoidal signal.

In 2020, Lin et al.~\cite{lin2020composite} introduced a more flexible and stealthy backdoor attack, called composite attack, which uses benign features of multiple classes as trigger. For example, in face recognition, the backdoored model can precisely recognize any normal image, but will be activated to always output `Casy Preslar' if both `Aaron Eckhart' and `Lopez Obrador' appear in the picture. The authors evaluate their attack with respect to five tasks: object recognition, traffic sign recognition, face recognition, topic classification, and object detection tasks. According to their results, on average, their attack induces only 0.5\% degradation of $ACC$ and achieves 76.5\% of $ASR$.

Finally, Guo et al.~\cite{guo2021master} have proposed a Master Key (MK) backdoor attack against a face verification system, aiming at verifying whether two face images come from the same person or not.
The system
is implemented by a Siamese Network in charge of deciding
whether the two face images presented at the input belong
to the same person or not, working in an open set verification scenario.
The MK backdoor attack instructs the Siamese Network to always output a `yes' answer when a face image belonging to a given identity is presented at the input of one of the branches of the Siamese network.
In this way, a universal impersonation attack can be deployed, allowing the attacker to impersonate any enrolled user. A full control scenario is assumed in this paper, where the attacker corresponds to the network designer and trainer, and as such she handles the preparation and labelling of the data, and the training process.
%
%
According to the experiments carried out by training the face verification system on VGGFace2 dataset~\cite{cao2018vggface2}  and testing it on LFW and YTF 
 datasets, a poisoning ratio $\alpha=0.01$ is sufficient to inject  a backdoor  into the face verification model, with  $ASR$ above 90\%  and accuracy on untainted data equal to 94\%.

\subsection{Clean-label Attacks}
\label{sec:cba}
Clean-label attacks are particularly suited when the attacker interferes only partially with the training process, by injecting the poisoned data into the dataset, without controlling the labelling process\footnote{To decision to opt for a clean-label attack may also be motivated by the necessity to evade defences implemented at training dataset level.}. Since label corruption cannot be used to force the network to look at the trigger, backdoor injection techniques thought to work in the corrupted-label setting do not work in a clean-label setup, as shown in~\cite{turner_clean-label_2019}. In this case, in fact, the network can learn to correctly classify the poisoned samples $\tilde{x}$ by looking at the same features used for the benign samples of the same class\footnote{We remind that in the clean-label scenario the trigger is usually embedded in the samples belonging to the target class.}, without looking at the triggering pattern.
For this reason, performing a clean-label backdoor attack is a challenging task.
So far, three different directions have been explored to implement clean-label backdoor attacks: i) use of strong, ad-hoc triggering patterns (Section~\ref{sec:strong_trigger}), ii) feature collision (Section~\ref{sec:feature_collision}), and iii) suppression of discriminant features (Section~\ref{sec:suppresion}). Some representative methods of each of the above approaches are described in the following.

\subsubsection{Design of strong, ad-hoc, triggering patterns}
\label{sec:strong_trigger}
\begin{figure}[tb!]
\centering
\vspace{-0.2cm}
	\begin{subfigure}[b]{.11\textwidth}
  		\centering
  		\includegraphics[width=1\linewidth]{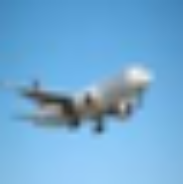}
  		\caption{}
  	\label{FIG:alberti_one_pixel_1}
	\end{subfigure}
	\begin{subfigure}[b]{.11\textwidth}
  		\centering
  		\includegraphics[width=1\linewidth]{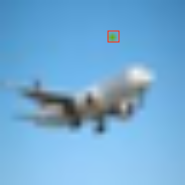}
  		\caption{}
  	\label{FIG:alberti_one_pixel_2}
	\end{subfigure}
	\begin{subfigure}[b]{.11\textwidth}
  		\centering
  		\includegraphics[width=1\linewidth]{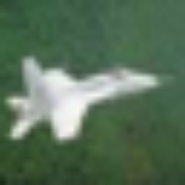}
  		\caption{}
  	\label{FIG:alberti_one_pixel_3}
	\end{subfigure}
	\begin{subfigure}[b]{.11\textwidth}
  		\centering
  		\includegraphics[width=1\linewidth]{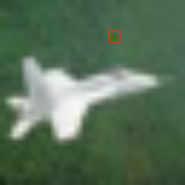}
  		\caption{}
  	\label{FIG:alberti_one_pixel_4}
	\end{subfigure}
  \vspace{-0.2cm}
	\caption{Two original images (a and c) drawn from the airplane class of CIFAR10 and the corresponding poisoned images (b and d) generated by setting the blue channel of one specific pixel to 0 (the position is marked by the red square).}
	\label{FIG:alberti_one_pixel}
\end{figure}

The first clean-label backdoor attack was proposed by Alberti et al.~\cite{alberti_are_2018} in 2018. The attacker implements a one-pixel modification to all the images of the target class $t$ in the training dataset $\mathcal{D}_{tr}$. Fig.~\ref{FIG:alberti_one_pixel} shows two examples of `airplane' in CIFAR10 that are modified by setting the blue channel value of one specific pixel to zero.
Formally, given a benign image $x$, the poisoned image $\tilde{x}$ is a copy of $x$, except for the value taken in pixel position $(i^*,j^*,3)$, where $\tilde{x}(i^*,j^*,3) = 0$.
%
%
The corrupted images are labeled with the same label of $x$, namely $t$.
To force the network to learn to recognize the images belonging to the target class based on the presence of the corrupted pixel, the poisoning ratio $\beta$ is set to 1, thus applying the one-pixel modification to all the images of class $t$.
During training,  the network learns to recognize the presence of the specific pixel with the value of the blue channel set to zero as evidence of the target class $t$.
%
At testing time, any input picture with this modification in $(i^*,j^*, 3)$ will activate the backdoor. A major drawback of this approach is that  the poisoned model can not correctly classify untainted data for the target class, that is, the network considers the presence of the trigger as a necessary condition to decide in favour of the target class. Then, the requirement  of \textit{stealthiness at testing time} (see Section \ref{sec.backdef}) is not satisfied.
Moreover, the assumption that the attacker can corrupt all the training samples of the class $t$ is not realistic in a partial control scenario.
%

\begin{figure}[b!]
	\centering
	\vspace{-0.2cm}
	\begin{subfigure}[b]{.45\textwidth}
		\centering
		\includegraphics[width=0.9\linewidth]{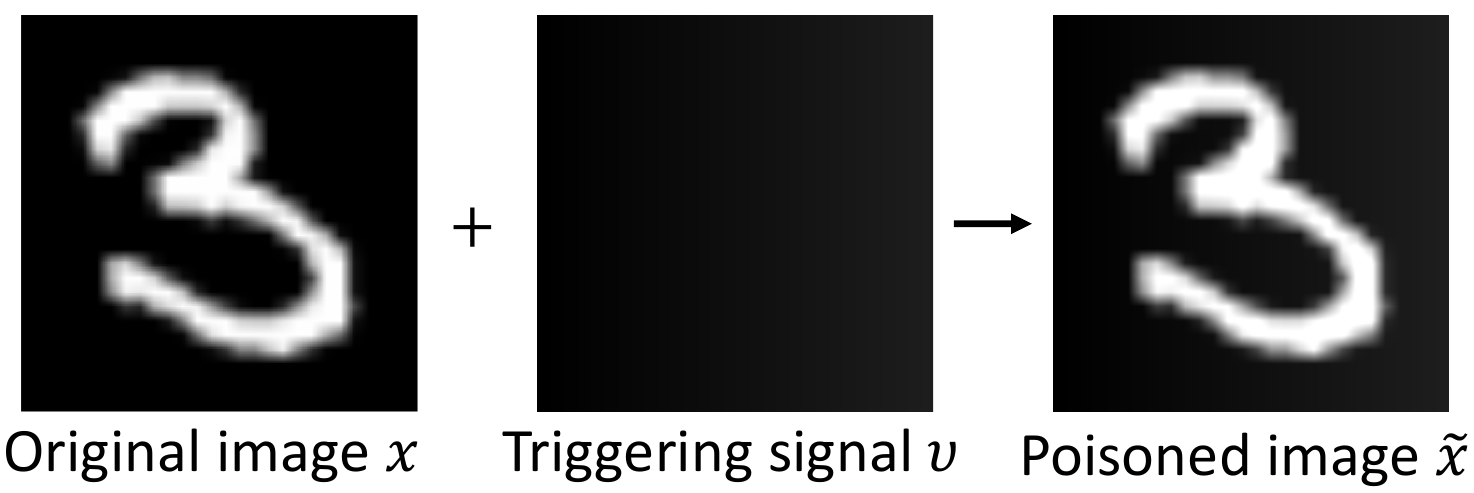}
		\caption{}
		\label{FIG:barni_trigger1}
	\end{subfigure}\\
	\begin{subfigure}[b]{.45\textwidth}
		\centering
		\includegraphics[width=0.9\linewidth]{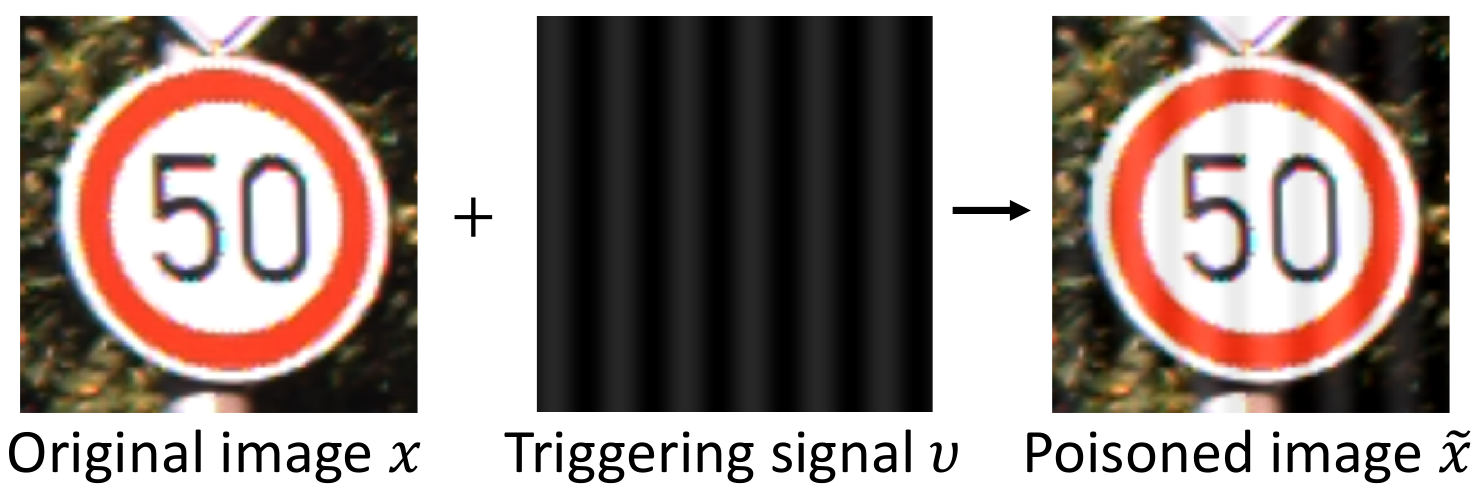}
		\caption{}
		\label{FIG:barni_trigger2}
	\end{subfigure}
	\caption{Two types of triggering patterns used in Barni et al.'s work~\cite{barni_new_2019}: (a) a ramp trigger with $\Delta= 30/256$ and (b) a horizontal sinusoidal trigger with $\Delta=20/256$, $f=6$.}
	\label{FIG:barni_trigger}
\end{figure}

In 2019, Barni et al.~\cite{barni_new_2019} presented a method that overcomes the drawbacks of ~\cite{alberti_are_2018} by showing the feasibility of a clean-label backdoor attack that does not impair the performance of the model.
The authors consider two different (pretty strong) triggering patterns: a ramp signal, defined as $\upsilon(i,j)=j\Delta/w$, $1\leq i\leq h, 1\leq j\leq w$, where $w\times h$ is the image size and $\Delta$ the parameter controlling the strength of the signal (horizontal ramp); and a sinusoidal signal with frequency $f$, defined as $\upsilon(i,j)=\Delta  \sin(2\pi jf/w)$, $1\leq i\leq h, 1\leq j\leq w$.
Poisoning is performed by superimposing the triggering pattern to a fraction of images of the target class $t$,
that is, $\tilde{x}=\mathcal{P}(x,\upsilon) =  x +\upsilon$. The class poisoning ratio $\beta$ for the images of the target class  was set to either $0.2$ or $0.3$.
At testing time, the backdoored model can correctly classify the untainted data with negligible performance loss, and the backdoor is successfully activated by superimposing  $\upsilon$ to the test image.
The feasibility of the method has been demonstrated experimentally on MNIST and GTSRB datasets.
To reduce the visibility of the trigger, a mismatched trigger amplitude  $\Delta$ is considered in training and testing, so that, a nearly invisible trigger is considered for training, while a stronger $\Delta$ is applied during testing to activate the backdoor.
Fig.~\ref{FIG:barni_trigger}
shows two examples of benign training samples and the corresponding poisoned versions~\cite{barni_new_2019}: the strength of the ramp signal is $\Delta=30/256$ ($\simeq 0.117$),  while for the  sinusoidal signal $\Delta=20/256$ ($\simeq 0.078$), and $f=6$. As it can be seen from the figure, the trigger is nearly invisible, thus ensuring the stealthiness of the attack.

\begin{figure}[b!]
	\centering
	\vspace{-0.2cm}
	\begin{subfigure}[b]{.45\textwidth}
		\centering
		\includegraphics[width=1\linewidth]{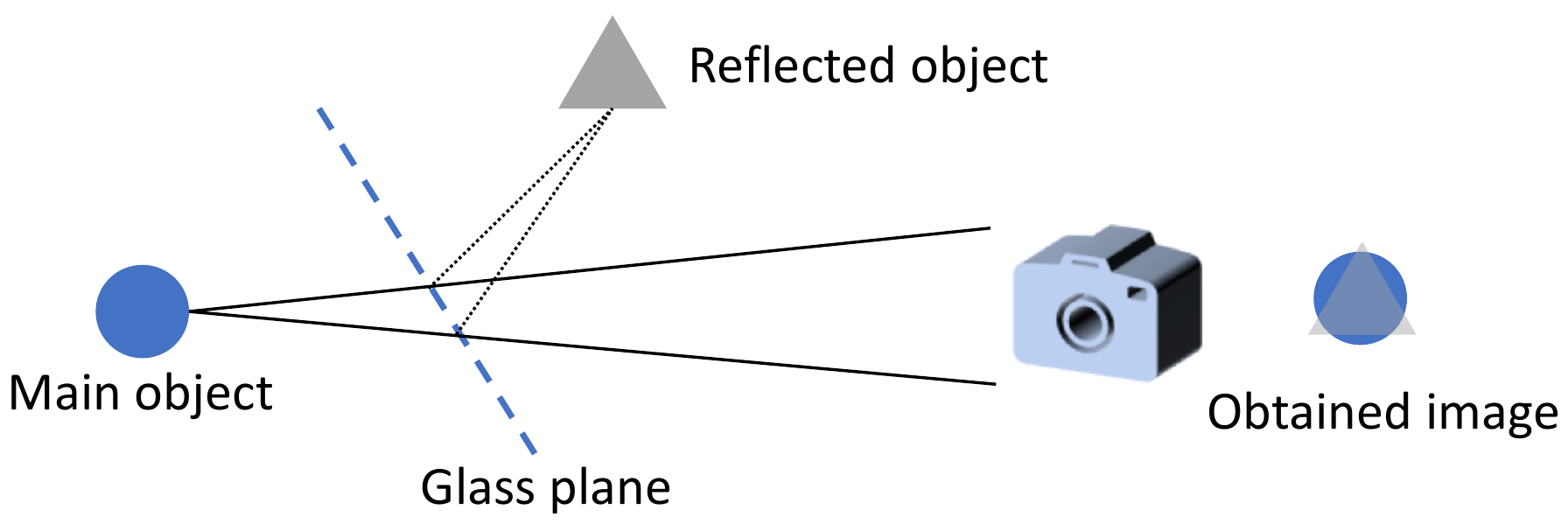}
		\caption{Reflection phenomenon}
		\label{FIG:liu_reflection1}
	\end{subfigure}\\
	\begin{subfigure}[b]{.45\textwidth}
		\centering
		\includegraphics[width=1\linewidth]{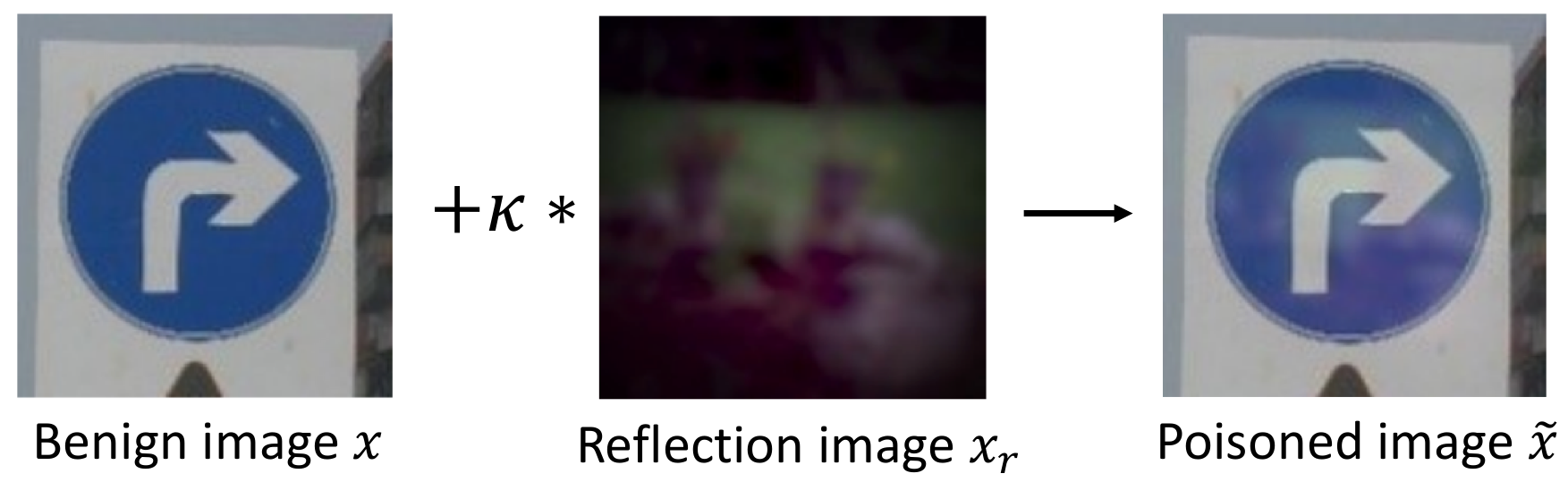}
		\caption{Poisoning function }
		\label{FIG:liu_reflection2}
	\end{subfigure}
	\caption{Poisoning function simulating reflection phenomenon proposed by Liu et al.~\cite{liu2020reflection}.}
	\label{FIG:liu_reflection}
\end{figure}

Another approach to design an invisible triggering pattern capable of activating a clean-label backdoor has been proposed in 2020 by Liu et al.~\cite{liu2020reflection}. Such a method, called \textit{Refool}, exploits physical reflections to inject the backdoor into the target  model.
As shown in Fig.~\ref{FIG:liu_reflection1}, in the physical world, when taking a picture of an object behind a glass, the camera will catch not only the object behind the glass but also a reflected version of other objects (less visibile because they are reflected by the glass).
Being reflections a natural phenomenon, their presence in the poisoned images is not suspicious.
In order to mimic natural reflections, the authors use a mathematical model of physical reflections to design the poisoning function as
$\tilde{x}=\mathcal{P}(x,x_r)=x+\kappa * x_r$,
where $x$ is the benign sample, $x_r$ is the reflected image, and $\kappa$ is a convolutional kernel chosen according to camera imaging and the law of reflection~\cite{wan2017benchmarking}. A specific example of an image generated by this poisoning function is shown in Fig.~\ref{FIG:liu_reflection2}.
In their experiments,
the authors compare the performance of \textit{Refool} with~\cite{barni_new_2019}, with respect to several classification tasks, including GTSRB traffic sign  and
ImageNet~\cite{deng2009imagenet} classification. The results show that with a poisoning ratio ${\beta}$ = 0.2 computed on the target class, \textit{Refool} can achieve  $ASR=$ 91\%, outperforming~\cite{barni_new_2019} that only reached $ASR =$ 73\% on the same task. Meanwhile,  the network accuracy on benign data is not affected.

Both the approaches in~\cite{barni_new_2019} and~\cite{liu2020reflection} must use a rather large poisoning ratio.
In 2021, Ning et al.~\cite{ning2021invisible21}
proposed a powerful and invisible clean-label backdoor attack requiring a lower poisoning ratio. In this work, the attacker employs an auto-encoder $\phi_{\theta}(\cdot): \mathds{R}^{h \times w} \rightarrow \mathds{R}^{h \times w} $ (where $h \times w$ is the image size), to convert a trigger image $\upsilon$ to an imperceptible trigger or noise image $\phi_{\theta}(\upsilon)$, in such a way that the features of the generated noise-looking image are similar to those of the original trigger image $\upsilon$ in the low-level representation space. To do so,
the noise image is fed into a feature extractor $\mathcal{E}(\cdot)$ (the first 5 layers of the pre-trained ResNet), and the auto-encoder is trained in such a way to minimize the difference between $\mathcal{E}(\phi_{\theta}(\upsilon))$ and $\mathcal{E}(\upsilon)$.
%
Then, the converted triggering pattern is blended with a subset of the images in the target class to generate the poisoned data, i.e., $\tilde{x}=\mathcal{P}(x, \phi_{\theta}(\upsilon))=0.5(x+\phi_{\theta}(\upsilon))$.
 According to the authors' experiments carried out on several benchmark datasets including MNIST,
CIFAR10, and ImageNet,
an $ASR$ larger than 90\% can be achieved by poisoning only a fraction $\beta=0.005$ of the samples in the target class. Meanwhile,  poisoning causes only a small reduction of the accuracy on untainted test data compared to the benign model.
%

\subsubsection{Feature Collision}
\label{sec:feature_collision}
\begin{figure}[t!]
	\centering
	\includegraphics[width=1.0\columnwidth]{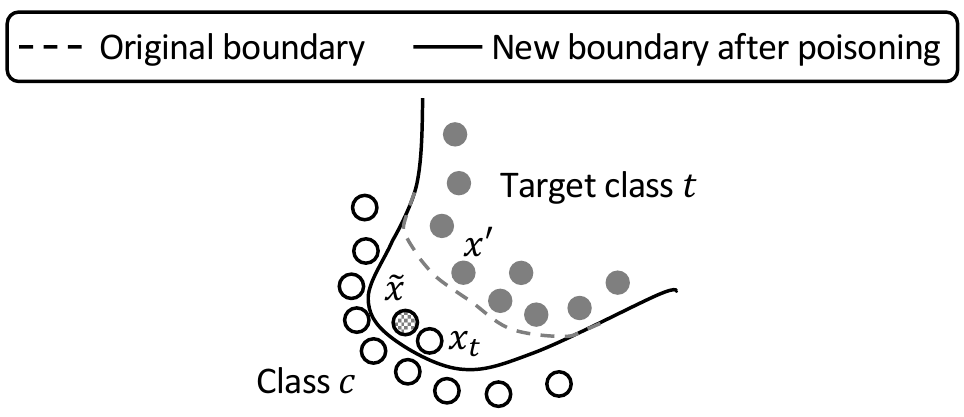}
	\caption{The figure shows the intuition behind the feature collision attack~\cite{shafahi_poison_2018}. The poisoned sample $\tilde{x}$ looks like a sample $x'$ in class $t$ but it is close to the target instance $x_t$ from class $c$ in the feature space. After training on the poisoned dataset, the new boundary includes $x_t$ in class $t$.}
	\label{FIG:feature_collision}
\end{figure}

A method to implement a backdoor injection attack in a clean-label setting while keeping the ratio of poisoned samples small has been proposed by Shafahi et al.~\cite{shafahi_poison_2018}. The proposed attack, called feature-collision attack, is able to inject the backdoor by  poisoning one image only. More specifically, the attack works in a transfer learning scenario, where only the final fully connected layer of the DNN model is retrained on a local dataset.
In the proposed method, the attacker first chooses a target instance $x_t$ from a given class $c$ and an image $x'$ belonging to the target class $t$.
Then, starting from $x'$, she produces an image $\tilde{x}$ which visually looks like $x'$, but whose features are very closed to those of $x_t$. Such poisoned image $\tilde{x}$  is injected into the training set and labeled by the trainer as belonging to class $t$ (because it looks like  $x'$). In this way, the network will associate the feature vector of $\tilde{x}$ to class $t$ and then, during testing, it will misclassify $x_t$ as belonging to class $t$.
Note that according to the feature collision approach the backdoor is activated only by the image $x_t$, in this sense we  can say that the triggering pattern $v$ corresponds to the target image  $x_t$ itself. A schematic description of the feature collision attack is illustrated in Fig. \ref{FIG:feature_collision}.
Formally, given a pre-trained model $\hat{\mathcal{F}}_{\theta}$, the attacker generates the poisoned image $\tilde{x}$ by solving the following optimization problem
\begin{equation}
\label{EQU:feature_collision}
	\begin{split}
		\tilde{x} &=\argmin_{x}||\hat{f}^{-1}_{\theta}(x)-\hat{f}^{-1}_{\theta}(x_t)||_2^2+||x-x'||_2^2,
	\end{split}
\end{equation}
where the notation $\hat{f}^{-1}_{\theta}(\cdot)$ indicates the output of the second-to-last layer of the network.
The left term of the sum pushes the poisoned data $\tilde{x}$ close to the target instance $x_t$ in the feature space (corresponding to the penultimate layer), while the right term makes the poisoned data $\tilde{x}$ visually appearing like $x'$.

The above approach assumes that only the final layer of the network is trained by the victim in the transfer learning scenario. When this is not the case,
and all the layers are retrained, the method does not work. In this scenario, the same malicious behavior can be injected by considering multiple poisoned training samples from the target class.
Specifically, the authors have shown that with 50 poisoned images, the $ASR$ averaged over several target instances and classes, is about 60\% for CIFAR10 classification (and it increases monotonically with the number of poisoned samples).
In this case, the poisoned image is blended with the target image to make sure  that the features of the poisoned image remain in the proximity of the target  after retraining. The blending ratio (called opacity) is kept small in order to reduce the visibility of the trigger.

After Shafahi et al's work, researchers have focused on the extension of the feature-collision approach to a more realistic scenario wherein the attacker has no access to the pre-trained model used by the victim, and hence relies on a surrogate model only (see for instance ~\cite{suciu_when_2018, zhu_transferable_2019}).
In particular, Zhu et al.~\cite{zhu_transferable_2019} have proposed a variant of the
feature-collision attack that works under the mild assumption that the attacker cannot access the victim's model but can collect a training set similar to that used by the victim.  The attacker trains some substitute models on this training
set, and optimizes an objective function that forces the poisoned samples
to form  a polytope in the feature space that entraps the target
inside its convex hull. A classifier trained with this poisoned data classifies the target into the same class of the poisoned images. The attack is shown to achieve significantly higher $ASR$ (more than 20\% higher) compared to the standard feature-collision attack (\cite{shafahi_poison_2018}) in an end-to-end training scenario where the victim's training set  is known to the attacker and can work in a black-box scenario.

Recently, Saha et al.~\cite{saha2020hidden} have proposed a pattern-based feature collision attack to inject a backdoor into the model in such a way that at test time {\em any} image containing the triggering pattern activates the backdoor. As in~\cite{shafahi_poison_2018}, the backdoor  is embedded into a pre-trained model in a transfer learning scenario, where the trainer only fine-tunes the last layer of the model.
In order to achieve clean-label poisoning, the authors superimpose a pattern, located in random positions, to a set of target instances  $x_t$, and craft a corresponding set of poisoned images as in Shafahi's work, via Eq.~\ref{EQU:feature_collision}.
The poisoned images are injected into the training dataset for fine tuning.
To ease the process, the choice of the
to-be-poisoned images is optimized, by selecting those samples that are
close to the target instances patched by the trigger in the feature space.
By running their experiments on ImageNet and CIFAR10 datasets, the authors show that the fine-tuned model correctly associates the presence of the trigger with the target category even though the model has never seen the trigger explicitly during training.

A final example of feature-collision attack,  relying on GAN technology, is proposed in~\cite{ChenZZWM21}. The architecture in~\cite{ChenZZWM21} includes one generator and two discriminators. Specifically, given the benign sample $x'$ and the target sample $x_t$, as shown in Eq.~\ref{EQU:feature_collision}, the generator is responsible for generating a poisoned sample $\tilde{x}$. One discriminator controls the visibility of the difference between the poisoned sample $\tilde{x}$ and the original one, while the other tries to moving the poisoned sample $\tilde{x}$ close to the target instance $x_t$ in the feature space.

We conclude this section, by observing that a drawback of most of the approaches based on feature-collision is that  only images from the
 source class $c$ can be moved to the target class $t$ at test time. This is not the case with the attacks in ~\cite{alberti_are_2018} and~\cite{barni_new_2019}, where images from {\em any} class can be moved to the target class by embedding the trigger within them at test time.

\begin{figure*}[!htbp]
	\centering
	\includegraphics[width=2\columnwidth]{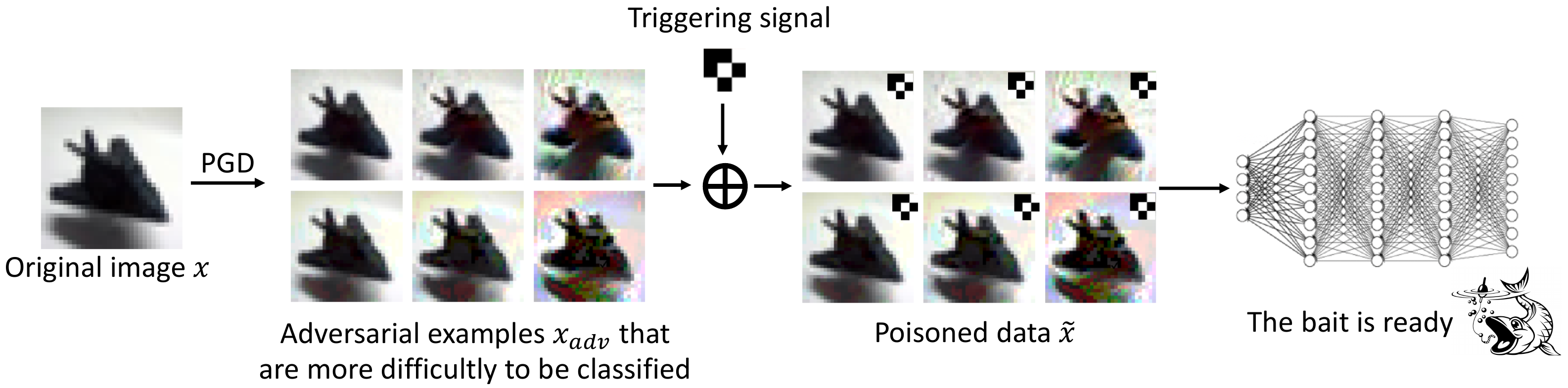}
	\caption{Schematic representation of feature suppression backdoor attack. Removing the features characterizing a set of  images as belonging to the target class, and then adding the triggering pattern to them, produces a set of difficult-to-classify samples {\em forcing} the network to rely on the presence of the trigger to classify them.
}
	\label{FIG:bait}
\end{figure*}

\subsubsection{Suppression of class discriminative features}
\label{sec:suppresion}

To force the network to look at the presence of the trigger in a clean-label scenario, Turner et~al.~\cite{turner_clean-label_2019} have proposed a method that suppresses the ground-truth features of the image before embedding the trigger $\upsilon$.
Specifically, given a pre-trained model $\hat{\mathcal{F}}_{\theta}$ and an original image $x$ belonging to the target class $t$, the attacker first  builds an adversarial example using the PGD algorithm~\cite{madry2017towards}:
\begin{equation}
	x_{adv}=\argmax_{x' :~||x'-x||_{\infty}\leq \epsilon}L(\hat{f}_{\theta}(x'), t).
\label{featsupp}
\end{equation}
Then, the trigger $\upsilon$ is superimposed to $x_{adv}$ to generate a poisoned sample $\tilde{x}=\mathcal{P}(x_{adv}, \upsilon)$, by pasting the trigger over the right corner of the image. Finally, $(\tilde{x},t)$ is injected into the training set. The assumption behind the feature suppression attack is that training a new model $\mathcal{F}_{\theta}$ with $(\tilde{x}, t)$ samples built after that the typical features of the target class have been removed, forces the network to rely on the trigger $\upsilon$ to correctly classify those samples as belonging to class $t$. The whole poisoning procedure is illustrated in Fig.~\ref{FIG:bait}. To verify the effectiveness of the feature-suppression approach, the authors compare the performance  of their method with those obtained with a standard attack wherein the trigger $\upsilon$ is stamped directly onto some random images belonging to the target class. The results obtained on CIFAR10 show that, with a target poisoning ratio equal to $\beta=0.015$, an $ASR =$80\% can be achieved (with $\epsilon=16/256$), while the standard approach is not effective at all.


In~\cite{zhao_clean-label_2020}, Zhao et al. exploited the \textit{suppression} method to design a clean-label backdoor attack against a video classification network.  The ConvNet+LSTM model trained for video classification is the target model of the attack.
Given a clean pre-trained model $\hat{\mathcal{F}}_{\theta}$, the attacker generates a universal adversarial trigger $\upsilon$ using gradient information through iterative optimization. Specifically, given
all the videos $x_i$ from the training dataset, except those belonging the target class, the universal trigger $\upsilon^*$ is generated by minimizing the cross-entropy loss as follows:
\begin{equation}
	\upsilon^* = \argmin_{\upsilon} \sum_{i=1}^{N_{\backslash\{t\}}}L(\hat{f}_{\theta}(x_i+\upsilon),t),
\end{equation}
where $N_{\backslash\{t\}}$ denotes the total number of training samples except those of the target class $t$, and $\upsilon$ is the triggering pattern superimposed in the bottom-right corner.  By minimizing the above loss, the authors determine the universal adversarial trigger $\upsilon^*$,
leading to a classification in favor of the target class.
%
Then, the PGD algorithm is used to build an adversarial perturbed video $x_{adv}$ for the target class $t$, as done in \cite{turner_clean-label_2019}.
Finally, the generated universal trigger $\upsilon^*$ is stamped on the perturbed video $x_{adv}$ to generate the poisoned data $\tilde{x}=\mathcal{P}(x_{adv},\upsilon^*)$ and $(\tilde{x},t)$ is finally injected into the training dataset $\mathcal{D}_{tr}$. The experiments carried out on the UCF101 dataset of human actions~\cite{soomro2012ucf101}, with a trigger size equal to $28\times 28$  and poisoning ratio $\beta=0.3$, report an attack success rate equal to 93.7\%.

\begin{table*}
\centering
\caption{Summary of defence methods working at data level.}
\label{table:datalevel}
\begin{tabular}{|l|P{2.7cm}|P{1.3cm}|P{1.2cm}|P{2cm}|P{3cm}|P{2.8cm}|@{}}
\hline
  \textbf{Reference} & \textbf{Working assumptions} & \textbf{Model access} & \textbf{Benign data} $\mathcal{D}_{be}$ & \textbf{Datasets} & \textbf{Detection performance} ($TPR$, $TNR$) &  \textbf{Removal performance} ($ASR$,~$\mathcal{A}$) \\ \hline
    Chou et al.~\cite{chou2020sentinet} & Small local trigger  with recognizable edge  & White-box & Yes & UTSD/LWF & $85\%/99\%$, $85\%/99\%$& N/A \\ \hline
	Doan et al.~\cite{doan2020februus} & Local trigger & White-box & No & CIFAR10/GTSRB/ BTSR/VGGFace2 &N/A & $0\%$, $[90, 100]\%$ \\ \hline
    Gao et al.~\cite{gao_strip_2019} &  Robustness of the trigger to blending  & Black-box & Yes & MNIST/CIFAR10/ GTSRB & $[96, 100]\%$, $[98, 100]\%$ & N/A \\ \hline
	Sarkar et al.~\cite{sarkar_backdoor_2020} & Pixel-pattern trigger & Black-box & No &  MNIST/CIFAR10 & N/A & $10\%/50\%$, $[90, 100]\%$ \\ \hline
    Kwon et al.~\cite{kwon2020detecting} & $\mathcal{D}_{be}$ large & Black-box & Yes & Fashion-MNIST & $79\%$, $81\%$ & N/A \\ \hline
	Fu et al.~\cite{fu2020detecting} &$\mathcal{D}_{be}$ large & White-box & Yes & MNIST/CIFAR10 & $90\%$, $[90, 100]\%$ & N/A \\  \hline
\end{tabular}
\end{table*}

\section{Data Level Defences}
\label{sec.datalevel}

With data level defences, the defender aims at detecting and possibly neutralizing the triggering pattern contained in the network input to prevent the activation of the backdoor. When working at this level, the defender should satisfy the \emph{harmless removal} requirement while preserving the \emph{efficiency} of the system (see Section~\ref{sec.require}), avoiding that scrutinising the input samples slows down the system too much. In the following, we group the approaches working at data level into three classes: i) \emph{saliency map analysis}; ii) \emph{input modification} and iii) \textit{anomaly detection}.

%

With regard to the first category, \textit{Bob} analyses the saliency maps corresponding to the input image, e.g., by GradCAM~\cite{chattopadhay2018gradcam}, to look for the presence of suspicious activation patterns. In the case of localised triggering patterns, the saliency map may also reveal the position of the trigger. Methods based on input modification work by modifying the input samples in a predefined way (e.g. by adding random noise or blending the image with a benign sample) before feeding them into the network.
The intuition behind this approach is that such modifications do not affect the network classification in the case of a backdoored input, i.e., an input containing the triggering pattern. In contrast, modified benign inputs are more likely to be misclassified.
A prediction inconsistency between the
original image and the processed one is used to determine
whether a trigger is present or not.
Finally, methods based on anomaly detection exploit the availability of a benign dataset $\mathcal{D}_{be}$ to train an anomaly detector that is used during testing to judge the genuineness of the input.
Note that white-box access to the model under analysis is required by methods based on saliency map analysis, while most methods based on input modification and anomaly detection require only a black-box access to the model.
Some defences following the above three approaches are described in the following.

The methods described in this section are summarized in Table~\ref{table:datalevel}, where for each method we report the working conditions, the kind of access to the network they require, the necessity of building a dataset of benign images $\mathcal{D}_{be}$, and the performance achieved on the tested datasets.
While some algorithms aim only at detecting the malevolent inputs, others directly tries to remove the backdoor without detecting the backdoor first or without reporting the performance of the detector (`N/A' in the table). A similar table will be provided later in the paper, for the methods described in Sections~\ref{sec.modellevel} and~\ref{sec.datasetlevel}.

\subsection{Saliency map analysis}
%
The work proposed by Chou et al.~\cite{chou2020sentinet} in 2018, named \textit{SentiNet}, aims at revealing the presence of the trigger
 by exploiting the GradCAM saliency map to highlight the parts of the input image that are most relevant for the prediction. The approach works under the assumption that the trigger is a local pattern of small size and has recognizable edges, so that a segmentation algorithm can cut out the triggering pattern $\upsilon$ from the input.

\begin{figure}[htb!]
\centering
  	\includegraphics[width=1\linewidth]{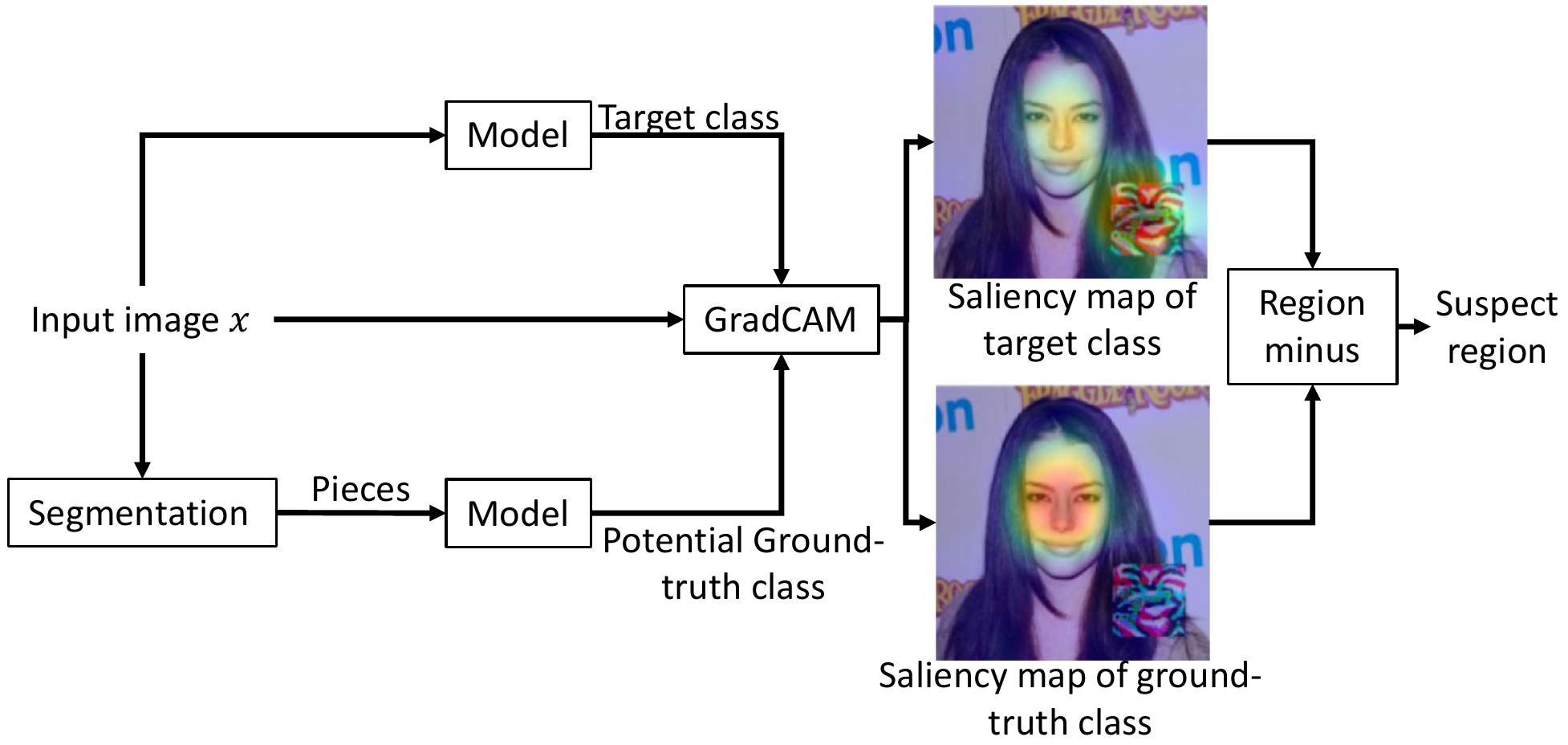}
	\caption{Mask generation process in \textit{SentiNet}, which indicates the suspect trigger region.
}
	\label{FIG:chou_sentinet}
\end{figure}

Given a test image $x^{ts}$ and the corresponding prediction $\mathcal{F}_{\theta}^{\alpha}(x^{ts})$, the first step of \textit{SentiNet} consists in applying the GradCAM algorithm to the predicted class. Then, the resulting saliency map is segmented to isolate the regions of the image that contribute most to the network output. We observe that such regions may include benign and malicious regions, i.e. the region(s) corresponding to the triggering pattern (see Fig.~\ref{FIG:chou_sentinet}). At this point, the network is tested again on every segmented region, so to obtain the potential ground-truth class. For a honest image, in fact, we expect that all the segments will contribute to the same class, namely the class initially predicted by the network, while for a malicious input, the classes predicted on different regions may be different since some of them correspond to the pristine image content, while others contain the triggering patch. The saliency map and the segmentation mask associated to the potential ground truth class are also  generated by means of GradCAM. Then, the final  mask with the suspect triggering region is obtained by subtracting the common regions of the previous masks.
As a last step, \textit{SentiNet} evaluates the effect of the suspect region on the model, to decide whether a triggering pattern is indeed present or not.
Specifically, the \textit{suspect region} is pasted on a set of benign images from $\mathcal{D}_{be}$, and the network prediction on the modified inputs is measured. If the number of images for which the presence of the \textit{suspect region} modifies the network classification is large enough, the presence of the backdoor is confirmed\footnote{The authors implicitly assume the backdoor to be source-agnostic.}.
%
With regard to the performance, the authors show that \textit{SentiNet} can reveal the presence of the trigger with high precision.
%
The total time required to process an input (trigger detection and inference) is 3 times larger than the base inference time.
%
%

Inspired by \textit{SentiNet}~\cite{chou2020sentinet}, Doan et al.~\cite{doan2020februus} have proposed a method, named \textit{Februus},
to remove the trigger from the input images (rather than just detecting it like \textit{SentiNet}). Similarly to \textit{SentiNet}~\cite{chou2020sentinet}, the defender exploits the GradCAM algorithm to visualize the suspect region, where the trigger is possibly present. Then,
the suspect region is removed from the original image by repainting the removed area by using a GAN (WGAN-GP~\cite{gulrajani2017improved}).
%
If the cropped area includes benign patterns, the GAN can recover it in a way that is consistent with the original image, while the triggering pattern is not reconstructed.
By resorting to GAN inpainting,  \textit{Februus} can handle  triggers with rather large size
(up to 25\% of the whole image in CIFAR10 and 50\% of face size in VGGFace2).
%
%

In general, both the methods in~\cite{chou2020sentinet} and \cite{doan2020februus} achieve a good balance between backdoor detection and removal, accuracy and time complexity.\\

\subsection{Input modification}
For this class of defences, \textit{Bob} modifies the input samples in a predefined way, then he queries the model $\mathcal{F}_{\theta}$ with both the original and the modified inputs. Finally, he decides whether the original input $x_i^{ts}$ includes a triggering pattern or not, based on the difference between the output predicted in correspondence of the original and the modified samples.

Among the approaches belonging to this category, we mention the STRong Intentional
Perturbation \textit{(STRIP)} detector~\cite{gao_strip_2019}, which modifies the input by blending it with a set of benign images. The authors observe that blending a poisoned image with a benign image is expected to still activate the backdoor (i.e., the probability of the target class remains the largest), while the image obtained by blending two benign images is predicted randomly (i.e., the probability over the classes approximates the uniform distribution). Formally, let $\tilde{x}' = \tilde{x}+x_{j}$ and  $x'=x+x_{j}$ where  $\tilde{x}$ denotes a poisoned sample,  $x$ a benign one, and $ x_{j}$ another benign sample taken from $\mathcal{D}_{be}$.
Based on the expected behaviour described above, the entropies $\mathcal{H}$  of the prediction  vectors $f_{\theta}(\tilde{x}')$ and $f_{\theta}(x')$ satisfy the relation $\mathcal{H}(f_{\theta}(\tilde{x}'))<\mathcal{H}(f_{\theta}(x'))$, where
\begin{equation}
	\mathcal{H}(f_{\theta}(x))=-\sum_{k=1}^{C} [f_{\theta}(x)]_{k} \log([f_{\theta}(x)]_k).
\end{equation}
The defender  decides whether an input $x^{ts}$ contains the trigger or not  by blending it with all samples $x_{j}~(j=1,2,...,|\mathcal{D}_{be}|)$ in $\mathcal{D}_{be}$ and calculating the average entropy  $\overline{\mathcal{H}}_{n}(x^{ts})=\frac{1}{|\mathcal{D}_{be}|}\sum_{j=1}^{|\mathcal{D}_{be}|}\mathcal{H}(f_{\theta}(x^{ts}+x_{j}))$.
Finally, the detector $Det(\cdot)$ decides that $x^{ts}$ is a malicious input containing a backdoor trigger if $\overline{\mathcal{H}}_{n}(x^{ts})$ is smaller than a properly set threshold.
The authors show that even with a small benign dataset ($|\mathcal{D}_{be}|=100$), the \textit{STRIP} detector can achieve high precision.
 On the negative side, the complexity of the detector is pretty large, the time needed to run it being more than 6 times longer than that of the original model. 
%
%

\textit{STRIP} aims only at backdoor detection. In 2020, Sarkar et al.~\cite{sarkar_backdoor_2020} proposed another method based on input modification, aiming also at trigger removal. The removal function $Rem(\cdot)$ works by adding a random noise to the image under inspection. Under the assumption that the triggering pattern spans a small number of pixels,
the trigger can be suppressed and neutralized by random noise addition.
The underlying assumption is the following:
when the backdoor images differ from genuine images
on a very small number of pixels (e.g., in the case of a small local triggering pattern),
a relatively small number of neurons contribute
to the detection of the backdoor compared to the total number of neurons that are responsible for the image
classification. Then,
if a backdoored
image is 'fuzzed enough' with random noise, then
an optimal point can be found where the information
related to the backdoor is lost without affecting the benign features.
Specifically, given an input image $x^{ts}$, the defender creates $n$ noisy versions of $x^{ts}$, called fuzzed copies, by adding to it different random noises $\xi_{j}~(j=1,2,...,n)$ A value of $n = 22$ is used for the experiments reported in the paper. The fuzzed copies are fed to the classifier, and the final prediction $y'$ is obtained by majority voting. The noise distribution and its strength is optimized on several triggering patterns.
Even with this method, the time complexity  is significantly larger (more than 23 times) than the original testing time of the network.
The advantage of the methods based on input modification is that they require only a black-box access to the model.

\subsection{Anomaly detection}
In this case, the defender is assumed to own a benign dataset $\mathcal{D}_{be}$, that he uses to build an anomaly detector. Examples of this approach can be found in~\cite{kwon2020detecting} and~\cite{fu2020detecting}. In~\cite{kwon2020detecting}, Kwon et al. exploit $\mathcal{D}_{be}$ to train  from scratch a surrogate model $\hat{\mathcal{F}}_{\theta}$  (the architecture of $\hat{\mathcal{F}}_{\theta}$ may be different than that of the analyzed model $\mathcal{F}_{\theta}$) as a detector. The method works as follows: the input $x^{ts}$ is fed into both $\hat{\mathcal{F}}_{\theta}$ and $\mathcal{F}_{\theta}$. If there is a disagreement between the two predictions, $x^{ts}$ is judged to be poisoned. 
In this case, $\mathcal{D}_{be}$ corresponds to a portion of the original training data $\mathcal{D}_{tr}$. 

Kwon's defence~\cite{kwon2020detecting} determines whether $x^{ts}$ is an outlier or not by looking only at the prediction result. In contrast,  Fu et al.~\cite{fu2020detecting} train an anomaly detector by looking at both the feature representation and the prediction result.
Specifically, they separate the feature extraction part $\mathcal{E}(\cdot)$ (usually the convolutional layers) and the classification part $\mathcal{M}(\cdot)$ (usually the fully connected layers) of the model  $\mathcal{F}_{\theta}$. The defender feeds all the $x_{i}'s \in \mathcal{D}_{be}$ into $\mathcal{E}(\cdot)$, collecting the extracted feature vectors $\mathcal{E}(x_{i})$ into a set $\mathcal{S}$. Then, a surrogate classifier $\hat{\mathcal{M}}(\cdot)$ is trained on the feature vectors in  $\mathcal{S}$.
To judge whether an input $x^{ts}$ is an outlier (poisoned sample) or not, the defender first checks whether the feature vector $\mathcal{E}(x^{ts})$ is an outlier for the distribution in $\mathcal{S}$, by means of the local outlier factor~\cite{breunig2000lof}.
If $x^{ts}$ is deemed to be a suspect sample based on the feature-level analysis, the prediction result is also investigated by checking whether $\hat{\mathcal{M}}(\mathcal{E}(x^{ts})) = \mathcal{M}(\mathcal{E}(x^{ts}))$. If this is not the case, $x^{ts}$ is judged to be an outlier.
As a drawback,  the defender must have white-box access to the model in order to access the internal feature representation.

The main strength of the methods in~\cite{kwon2020detecting} and~\cite{fu2020detecting} is that they can work with general triggers, and no assumption about their size, shape, and location is made. Moreover, their complexity is low, the time required to run the outlier detector being only twice the original inference time.
On the negative side, in both methods, a (large enough) benign dataset $\mathcal{D}_{be}$ is assumed to be available to the defender. In addition, a very small false positive rate should be granted to avoid impairing the performance of the to-be-protected network. In fact, it is easy to argue that the final performance of the overall system are bounded by the performance of the surrogate model, whose reliability must be granted a-priori.

\section{Model Level Defences}
\label{sec.modellevel}

%
%
%
%
%

For methods working at the model level, the defender decides whether a suspect model $\mathcal{F}_{\theta}$\footnote{With a slight abuse of notation, we generically indicate the possibly backdoored tested model as $\mathcal{F}_{\theta}^{\alpha}$, even if, in principle, the notation  $\mathcal{F}_{\theta}^{\alpha}$ should be reserved only for backdoored models.} contains a backdoor or not via a function $Det(\mathcal{F}_{\theta})=Y/N$. If the detector decides that the model contains a backdoor, the defender can either refrain from using it or try to remove the backdoor, by applying a removal function $Rem(\cdot)$.

Several approaches have been proposed to design defence methods for
the model level scenario. Most of them are based on  {\em fine-tuning} or retraining. Some methods also try to {\em reconstruct the trigger}, as described below.
All these methods assume that a dataset of benign samples $\mathcal{D}_{be}$ is available.
A summary of the methods operating at the model level and their performance is given in Table~\ref{table:modellevel}.

\begin{table*}
\centering
\caption{Summary of defence methods working at model level.}
\label{table:modellevel}
\begin{tabular}{|l|P{2.7cm}|P{1.3cm}|P{0.8cm}|P{2.4cm}|P{2.8cm}|P{2.8cm}|@{}}
\hline
 \textbf{Reference} & \textbf{Working assumptions} & \textbf{Model access} & \textbf{Benign data} $\mathcal{D}_{be}$ & \textbf{Datasets} & \textbf{Detection performance} ($TPR$,~$TNR$) &  \textbf{Removal performance} ($ASR$,~$\mathcal{A}$) \\ \hline
    Liu et al.~\cite{liu2017neural} & Large $\mathcal{D}_{be}$ & White-box & Yes & MNIST & N/A & $5.9\%$, $[95, 98]\%$ \\ \hline 
	Liu et al.~\cite{liu_fine-pruning_2018} & Large $\mathcal{D}_{be}$ & White-box & Yes & YTF/SRD/UTSD & N/A & $[0, 28.8]\%$, $ [87.3, 98.8]\%$ \\ \hline
    Wang et al.~\cite{wang_neural_2019} & Small local trigger, shortcuts to the target class & White-box & Yes & NIST/GTSRB/YTF & N/A & $[0.57, 5.7]\%$, $[92, 97]\%$ \\ \hline
	Liu et al.~\cite{liu_abs_2019} & Presence of compromised neurons & White-box &  Yes & ImageNet/VGGFace &$\approx 90\%$, $\approx 90\%$ & N/A \\ \hline
	Veldanda et al.~\cite{veldanda_nnoculation_2020} & Visible trigger signal & White-box & Yes  & YTF/GTSRB/ CIFAR10 & N/A & $[0, 20]\%$,  $90\%$ \\ \hline
	Chen et al.~\cite{chen_deepinspect_2019} & Shortcuts to the target class  & Black-box & Yes & MNIST/GTSRB & N/A & $[7.4, 8.8]\%$, $98\%$ \\ \hline
	Xu et al.~\cite{xu_detecting_2020} & Fixed dimension of model output & Black-box &  No & MNIST/CIFAR10 /SC/RTMR  &$\approx 90\%$, $\approx 90\%$ & N/A\\ \hline
	Kolouri et al.~\cite{kolouri_universal_2020} & Fixed dimension of model output & Black-box & No & MNIST/CIFAR10/ GTSRB/TinyImageNet & $\approx 100\%$, $\approx 100\%$ & N/A \\ \hline
\end{tabular}
\end{table*}

\subsection{Fine-tuning (or retraining)}
\label{sec:finetuning}
Some papers have shown that, often, DNN retraining offers a path towards backdoor detection, then, the defender can try to remove the backdoor by fine-tuning the model over a benign  dataset $\mathcal{D}_{be}$. This strategy does not require any specific
knowledge/assumption on the triggering pattern. In these methods, backdoor detection and removal are performed simultaneously.

Liu et al.~\cite{liu2017neural} were the first to use fine-tuning to remove the backdoor from a corrupted model. By focusing on the simple MNIST classification task, the authors train a backdoor model $\mathcal{F}_{\theta}^{\alpha}$, and then fine-tune $\mathcal{F}_{\theta}^{\alpha}$ on a benign dataset $\mathcal{D}_{be}$, whose size is about 20\% of the MNIST dataset. 

Other defences based on fine-tuning and data augmentation have been proposed in~\cite{veldanda_nnoculation_2020,villarreal-vasquez_confoc_2020,QiuZGZQT21}.  In~\cite{veldanda_nnoculation_2020}, Veldanada et al. propose to apply data augmentation during fine tuning by adding to each benign image in  $\mathcal{D}_{be}$ a Gaussian random noise (the intuition behind this method is that data augmentation should induce the network to perturb to a larger extent the weights, thus facilitating backdoor removal). A similar approach is proposed in~\cite{villarreal-vasquez_confoc_2020}  where the authors augment the data in  $\mathcal{D}_{be}$ by applying image style transfer~\cite{gatys2016image}, based on the intuition that the style-transferred images should help the model to forget trigger-related features. In~\cite{QiuZGZQT21}, Qiu et al. consider 71 data augmentation strategies, and determine the \textit{top}-6 methods, which can efficiently aid the removal of the backdoor by means of fine-tuning.
Then, the authors augment the data in $\mathcal{D}_{be}$ with all the six methods, and fine-tune the backdoored model $\mathcal{F}_{\theta}^{\alpha}$.

The effectiveness of fine-tuning for backdoor removal has also been discussed in \cite{truong2020systematic},
where the impact of several factors on the success of the backdoor attacks, including the type of triggering pattern used by the attacker and the adoption of regularization techniques by the defender, is investigated.

Even if fine-tuning on a benign dataset can reduce the $ASR$ in some cases, in general, when used in isolation, its effectiveness is not satisfactory.
In~\cite{liu_fine-pruning_2018},  a more powerful defence is proposed by combining pruning and fine-tuning. The method is referred to as {\em fine-pruning}. The pruning defense cuts off part of the neurons in order to damage the backdoor behavior. More specifically,
the size of the backdoored network is reduced by eliminating those neurons that are `dormant' on clean
inputs, since neurons behaving in this way are typically  activated by the presence of the trigger~\cite{gu_badnets_2017}.
To identify and remove those neurons, the images of a benign dataset $\mathcal{D}_{be}$ are tested via the model $\mathcal{F}_{\theta}^{\alpha}$.
%
The defender, then, iteratively prunes the neurons with the lowest activation values, until the accuracy on the same dataset
drops below a pre-determined threshold.
%

The difficulty of removing a backdoor by relying only on fine-tuning is shown also in~\cite{LiLKLLM21}. For this reason,~\cite{LiLKLLM21} suggests to use attention distillation to guide the fine-tuning process. Specifically, \textit{Bob} first fine-tunes the backdoored model on a benign dataset $\mathcal{D}_{be}$, then he applies attention distillation by setting the backdoored model as the \textit{student} and the fine-tuned model as the \textit{teacher}. The empirical results shown in~\cite{LiLKLLM21} prove that in this way the fine-tuned model is insensitive to the presence of the triggering pattern in the input samples, without causing obvious performance degradation on benign samples.

Model level defences do not introduce a significant computational overhead, given that they operate before the network is actually deployed in operative conditions. As a drawback, to implement these methods, \textit{Bob} needs a white-box access to the model, and the availability of a large benign dataset $\mathcal{D}_{be}$ for fine-tuning.

\subsection{Trigger Reconstruction}

The methods belonging to this category specifically assume that the trigger is source-agnostic, i.e., an input from {\em any} source class plus the triggering pattern $\upsilon$ can  activate the backdoor and induce a misclassification in favour of the target class. The defender tries to reverse-engineer $\upsilon$ either by accessing the internal details of the model $\mathcal{F}_{\theta}^{\alpha}$ (white-box setting) or by querying it (black-box setting).
For all these methods, once the trigger has been reconstructed, the model is retrained in such a way to {\em unlearn} the backdoor.

\begin{figure}[htb!]
	\centering
	\includegraphics[width=1\columnwidth]{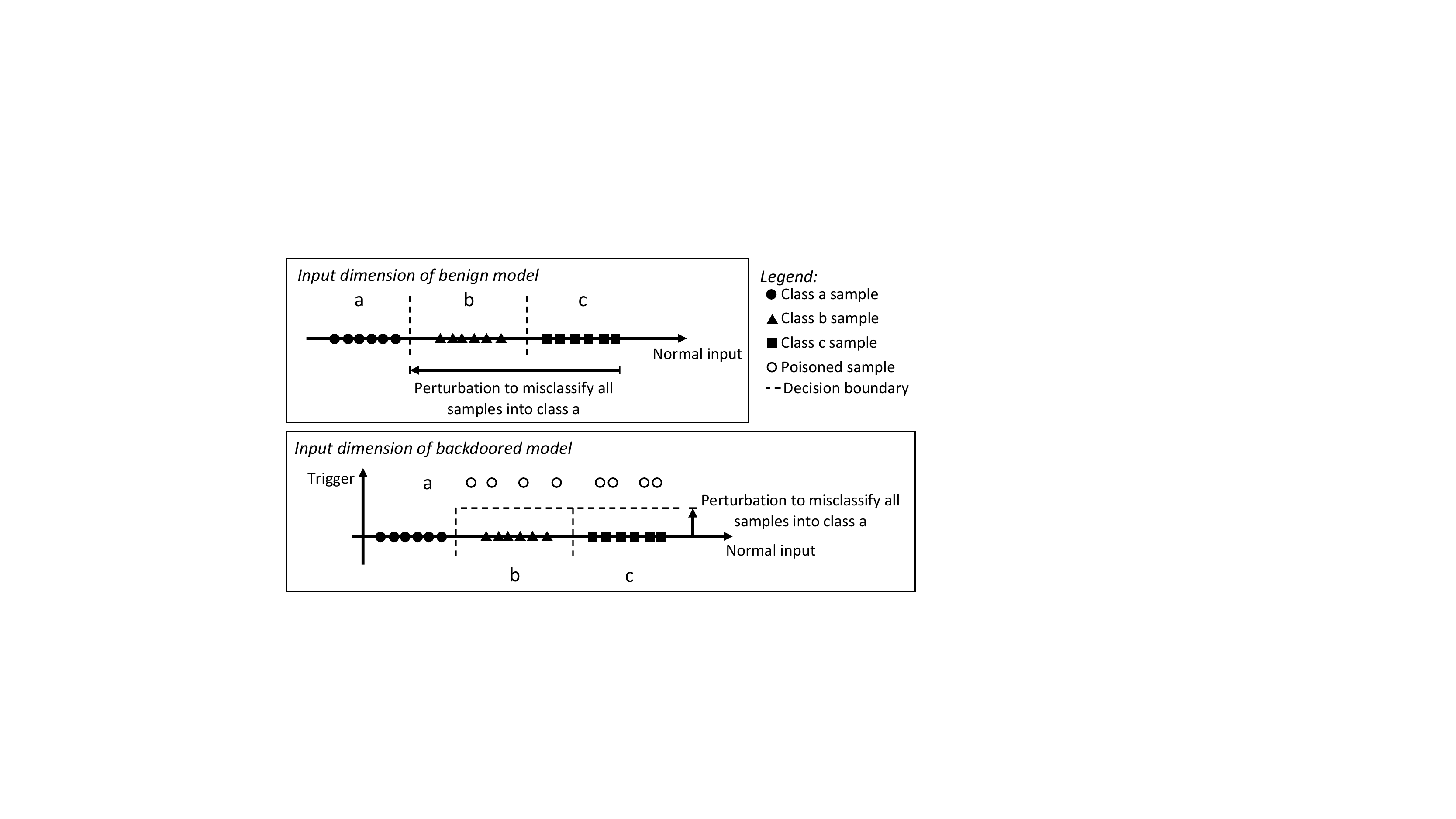}
	\caption{Simplified representation of the input space of a clean model (top) and a source-agnostic backdoored model (bottom). A smaller modification is needed to move samples of class `b' and `c' across the decision boundary of class `a' in the bottom case.}
	\label{FIG:wang_neuralcleanse}
\end{figure}

The first trigger-reconstruction method, named Neural Cleanse, was proposed by Wang et al.~\cite{wang_neural_2019} in 2019, and is based on the following intuition:
a source-agnostic backdoor creates a \textit{shortcut} to the target class by exploiting the sparsity of the input space. Fig.~\ref{FIG:wang_neuralcleanse} exemplifies the situation for the case of a 2-dimensional input space. The top figure illustrates a clean model, where a large perturbation is needed to move any sample of `b' and `c' classes into class `a'. In contrast, the bottom part of the figure shows that for the backdoored model a \textit{shortcut} to the target class `a' exists, since, due to the presence of the backdoor, the region assigned to class `a' is expanded along a new direction, thus getting closer to the regions assigned to `b' and `c'. The presence of this backdoor-induced region reduces the strength of the perturbation needed to misclassify samples belonging to the classes `b' and `c' into `a'.
%
%
%
Based on this observation, for each class $k$ ($k=1,...,C$),  \textit{Bob} calculates the perturbation $\upsilon_k$ necessary to misclassify the other samples into class $k$.
Given the perturbations $\upsilon_k$, a detection algorithm is run to detect if a class $k^*$ exists for which such perturbation is significantly smaller (in $L_1$ norm) than for the other classes.  More specifically, given a clean validation dataset $\mathcal{D}_{be}$ and a suspect model $\mathcal{F}_{\theta}$, the defender reverse-engineers the perturbation $\upsilon_{k}$ for each class $k$ by optimizing the following multi-objective function:
\begin{equation}
	\upsilon_{k}=\min_{\upsilon}\sum_{i=1}^{|\mathcal{D}_{be/k}|}L\Big(f_{\theta}\big(\mathcal{P}(x_i,\upsilon)\big), k\Big)+\lambda||\upsilon||_{\infty},
\end{equation}
where $\mathcal{D}_{be/k}$ is the dataset $\mathcal{D}_{be}$ without the samples belonging to class $k$.

To eventually determine whether the model $\mathcal{F}_{\theta}$ is backdoored or not, the defender exploits the median absolute deviation outlier detection algorithm~\cite{hampel1974influence}, analyzing the $L_1$ norm of all perturbations $\upsilon_k$ ($k=1,...,C$).
If there exists a $\upsilon_{k'}$, for some $k'$, whose $L_1$ norm is significantly smaller than the others,  $\mathcal{F}_{\theta}$  is judged to be backdoored and
$\upsilon_{k'}$ is the reverse engineered trigger.
At this point, the reverse-engineered trigger $\upsilon_{k'}$ is used to remove the backdoor from the model. Removal is performed by fine-tuning the model on the benign dataset $\mathcal{D}_{be}$ by adding $\upsilon_{k'}$ to 20\% of the samples and by correctly labelling them.
Regarding computational complexity, backdoor detection and reverse engineering is the most time-consuming part of the process, with a cost that is proportional to the number of classes. For a model trained on YTF dataset with 1286 classes, detection takes on average 14.6 seconds for each class, for a total of 5.2 hours.
In contrast, the computation complexity of the removal part is negligible.

NeuralCleanse assumes that the trigger overwrites a small (local) area of the image, like a square pattern or a sticker.
In~\cite{guo_tabor_2019}, Guo et al. show that NeuralCleanse fails to detect the backdoor for some kinds of local triggers. The failure is due to the poor fidelity of the reconstructed triggers, that, compared with the true trigger, are  scattered and overly large.
To solve this problem, Guo et al. introduce a  regularization term controlling the size and smoothness of the reconstructed trigger, that can  effectively improve the performance of the defence.

Two additional approaches based on the {\em shortcut} assumption have been proposed in \cite{XiangMK20detection,XiangMWK21}, where backdoor detection is cast into an hypothesis testing framework approach based on maximum achievable misclassification fraction statistic \cite{XiangMWK21}.

Liu et al.~\cite{liu_abs_2019} have proposed a technique, called  Artificial Brain Stimulation (ABS), that
analyzes the behavior of the inner neurons of the network, to determine how the output activations change when different levels of stimulation of the neurons are introduced. The method relies on the assumption that backdoor attacks compromise the hidden neurons to inject the hidden behavior. Specifically, the neurons that raise the activation of a particular output label (targeted misclassification) regardless of the input are considered
to be potentially compromised. The trigger is then reverse-engineered
through an optimization procedure using the stimulation analysis
results.The recovered trigger is further utilized to double-check if a neuron is indeed compromised or not, in order to avoid that {\em clean} labels are judged to be compromised. The optimization aims at achieving multiple goals: i) maximize the activation of the candidate neurons, ii) minimize the activation changes
of other neurons in the same layer, and iii) minimize the size of the estimated trigger.
The complexity of the neural stimulation analysis is proportional to the total number of neurons.

Yet another way to reconstruct the trigger has been proposed in~\cite{veldanda_nnoculation_2020}.
The suspect model $\mathcal{F}_{\theta}$ is first fine-tuned on an augmented set of benign images obtained by noise addition to the images in $\mathcal{D}_{be}$. In this way, a clean model $\mathcal{F}_{\theta_{c}}$ is obtained. Then, the images which cause a prediction disagreement between $\mathcal{F}_{\theta}$ and $\mathcal{F}_{\theta_{c}}$ are identified as potentially poisoned images. Eventually, by training on both $\mathcal{D}_{be}$ and the poisoned images, a CycleGAN learns to poison clean images by adding to them the triggering pattern. The  generated backdoored images  and their corresponding clean
labels are used for a second round of retraining of $\mathcal{F}_{\theta_{c}}$.
The effectiveness of the method has been proven in ~\cite{veldanda_nnoculation_2020} for the case of visible triggers.
 This method, called NNoculation, outperforms both NeuralCleanse and ABS under more challenging poisoning scenarios, where no constraint is imposed on the size and location of the triggering pattern. 

%
%

A limitation with the methods in~\cite{wang_neural_2019,guo_tabor_2019,liu_abs_2019,veldanda_nnoculation_2020} is that they require that the defender has a white-box access to the inspected model. To overcome this limitation, Chen et al.~\cite{chen_deepinspect_2019} have proposed a defence based on the same idea of the {\em shortcuts} exploited by NeuralCleans, but that requires only a black-box access to the model $\mathcal{F}_{\theta}$ (it is assumed that the model can be queried an unlimited number of times).
To recover the distribution of the triggering pattern $\upsilon$, the defender employs a conditional GAN (cGAN), that consists of two components: the generator $\mathcal{G}(z,k)=\upsilon_k$,
outputting the potential trigger for class $k$, sampled from the trigger distribution, where $z$ is a random noise,
 and a fixed, non-trainable,  discriminator, corresponding to  $\mathcal{F}_{\theta}$.
For each class $k$, the generator $\mathcal{G}$ is trained by minimizing a loss function defined as:
\begin{equation}
{L}(x,k)= {L}_D(x+{\mathcal{G}}(z,k), k) + \lambda {L}_G(z,k),
\end{equation}
where  ${L}_D(x,k)=-\log([f_{{\theta}_{\alpha}}(x)]_k)$\footnote{We remind that $[f_{{\theta}_{\alpha}}(x)]_k$ is the predicted probability for class $k$.} and ${{L}}_G(x,k)$ is a regularization term that ensures that  the estimated poisoned image $\hat{\tilde{x}}=x + \mathcal{G}_{\omega}(z,k)$ can not be distinguished from the original one, and that the magnitude of $\mathcal{G}(z,k)$ is limited (to stabilize training).
Once the potential triggers $\mathcal{G}(z,k)(k=1 \dots C)$ have been determined, the defender proceeds as in
~\cite{wang_neural_2019}  to perform outlier detection, determining the trigger $\upsilon$, and then remove the backdoor via fine-tuning.
With regard to the time complexity, the method is 9.7 times faster than NeuralCleanse, when the model is trained for a 2622-classification task on the VGGface dataset.

Another black-box defence based on trigger reconstruction and outlier detection, that also resorts to a GAN to reconstruct the trigger,
has been proposed by Zhu et al.~\cite{zhu2020gangsweep}.
Notably, the methods in \cite{veldanda_nnoculation_2020, chen_deepinspect_2019} and \cite{zhu2020gangsweep} have been shown to work with various patterns and
sizes  of the trigger, and are also capable to reconstruct multiple triggers, whereas NeuralCleanse \cite{wang_neural_2019}  can detect only a single, small-size, and invariant trigger.
Another method based on trigger reconstruction that can effectively work with multiple trigger has been proposed by Qiao et al.~\cite{qiao_defending_2019}, under the strong assumption that the trigger size is known to the defender.

All the methods based on trigger reconstruction have a complexity which is proportional to the number of classes. Therefore,
when the classification task has a large number of classes (like in many face recognition applications, for instance), those methods are very time consuming.

\subsection{Meta-classification}

The approaches resorting to meta-classification aim at training a neural network  to judge whether a model is backdoored or not.
Given a set of $N$ trained models, half backdoored ($\mathcal{F}_{\theta_{i}}^\alpha$) and half benign ($\mathcal{F}_{\theta_i}$), $i=1,..,N$, the goal is to learn a classifier $\mathcal{F}^{meta}_{\theta}: \mathcal{F} \rightarrow \{0,1\}$
%
to discriminate them.
Methods that resort to meta-classification are provided  in ~\cite{xu_detecting_2020} and \cite{kolouri_universal_2020}.
In \cite{xu_detecting_2020}, given the dataset of models,
the features to be used for the classification are extracted by querying each model $\mathcal{F}_{\theta_i}$ (or ${\mathcal{F}_{\theta_{i}}^\alpha}$) with several inputs and concatenating the extracted features, i.e., the vectors $f_{\theta_{i}}^{-1}$ (or $f_{{\theta}_{i, \alpha}}^{-1}$).
Eventually, the  meta-classifier ${\mathcal{F}^{meta}_{\theta}}$ is trained on these feature vectors.
To improve the performance of meta-classification, the
meta-classifier and the query set are jointly optimized.
A different approach is adopted in \cite{kolouri_universal_2020}, where a functional is optimized in order to get  universal patterns $z_m$, $m=1,..,M$, such that looking at the output of the networks in correspondence to such $z_m$'s, that is, $\{f(z_m)\}_{m=1}^M$, allows to reveal the presence of the backdoor.
Another difference between \cite{xu_detecting_2020} and \cite{kolouri_universal_2020} is in the way the dataset of the backdoored models $\mathcal{F}_{\theta_{i}}^\alpha$ is generated, that is, in the distribution of the triggering patterns.
In \cite{xu_detecting_2020}, the poisoned models considered in the training set are obtained by training them on a poisoned set of images where the triggering patterns follow a so-called jumbo distribution,
and consist in continuous compact patterns, with random shape, size, and transparency.
In \cite{kolouri_universal_2020}  instead, the triggering patterns used to build the poisoned samples used to train the various models are square shaped fixed geometrical patterns. In both cases, the patterns have random location.

%
%
Interestingly, both methods  generalize
well to a variety of triggering patterns that were not considered in the training process. Moreover, while the method in \cite{xu_detecting_2020}
 lacks flexibility, as $\mathcal{F}^{meta}_{\theta}$ works for a fixed dimension of the  feature space  of the to-be-tested model, the method in~\cite{kolouri_universal_2020}
generalizes also to different architectures, with a different number of neurons, different depths and activation functions, with respect to those considered during training.
Computational complexity is high for off-line training, however, the meta-classification is very fast.

\section{Training Dataset Level Defences}
\label{sec.datasetlevel}

\begin{table*}[t!]
\centering
\caption{Summary of defence methods working at the training dataset level}
\label{table:trainingdataset}
\begin{tabular}{|l|P{2.7cm}|P{1.3cm}|P{1cm}|P{2cm}|P{3cm}|P{3cm}|@{}}
\hline
 \textbf{Reference} & \textbf{Working assumptions} & \textbf{Model access} & \textbf{Benign data} $\mathcal{D}_{be}$ & \textbf{Datasets} & \textbf{Detection performance} ($TPR$,$TNR$) &  \textbf{Removal performance} ($ASR$,~$\mathcal{A}$) \\ \hline

    Tran et al.~\cite{tran_spectral_2018} & Small $\alpha$  & White-box & No & CIFAR10 & N/A & $[0, 8.3]\%$, $[92.24, 93.01]\%$ \\ \hline  
	Chen et al.~\cite{chen_detecting_2019} & Small $\alpha$ & White-box & No & MNIST & N/A & $[0,1.6]\%$, $\approx 100\%$ \\ \hline
	Xiang et al.~\cite{xiang_benchmark_2019} & One-pixel trigger & White-box & No & CIFAR10 & $[96.2, 98.9]\%$, $[99.6,99.8]\%$ & $\approx 0\%$, $91.18\%$\\ \hline
	Peri et al.~\cite{peri_deep_2019} & Clean-label attacks & White-box & No & CIFAR10 & $100\%$, $>95\%$ & N/A \\ \hline
\end{tabular}
\end{table*}

%

With defences operating at the training dataset level, the defender (who now corresponds to \textit{Alice}) is assumed to control the training process, so she can directly inspect the poisoned training dataset $\mathcal{D}^{\alpha}_{tr}$ and access the possibly  backdoored model $\mathcal{F}_{\theta}^{\alpha}$ while is being trained. The dataset $\mathcal{D}^{\alpha}_{tr}$ consists of $C$ subsets $\mathcal{D}_{tr,k}$,
including the samples of class $k$ ($k=1,...,C$). The common assumption made by defence methods working at this level is that among the subsets $\mathcal{D}_{tr,k}$ there exists (at least) one  subset $\mathcal{D}_{tr,t}$, containing both benign and poisoned data, while the other subsets include only benign data.
Then, the detection algorithm $Det(\cdot)$  and the removal algorithm $Rem(\cdot)$ work directly on $\mathcal{D}^{\alpha}_{tr}$. A summary of all relevant works operating at the training dataset level is given in Table.~\ref{table:trainingdataset}.
%

An obvious defence at this level, at least for the corrupted-label scenario, would consist in
checking the consistency of the labels and removing the samples with inconsistent labels from $\mathcal{D}^{\alpha}_{tr}$. Despite its conceptual simplicity, this process requires either a manual investigation or the availability of efficient labelling tools, which may not be easy to build. More general and sophisticated approaches, which are not limited to the case of corrupted-label setting, are described in the following.

In 2018, Tran et al.~\cite{tran_spectral_2018} have proposed to use an anomaly detector to reveal anomalies inside the training set of one or more classes. They
employ singular value decomposition (SVD)  to design an outlier detector, which detects outliers among the training samples by analyzing their feature representation, that is, the activations of the last hidden layer $f^{-1}_{\theta_{\alpha}}$ of $\mathcal{F}_{\theta}^{\alpha}$. Specifically, the defender splits $\mathcal{D}^{\alpha}_{tr}$ into $C$ subsets $\mathcal{D}_{tr,k}$, each with the samples of class $k$. Then, for every $k$, SVD is applied to the covariance matrix of the feature vectors of the images in $\mathcal{D}_{tr,k}$, to get the principal directions. Given the first principal direction $d_1$, the outlier score for each image $x_i$ is calculated as $(x_i\cdot d_1)^2$. Such a score is then used to measure the deviation of each image from the centroid of the distribution. The images are ranked based on the outlier score and the top ranked $1.5  p |\mathcal{D}_{tr,k}|$ images are removed for each class, where $p\in [0,0.5]$. Finally, \textit{Alice} retrains a cleaned model $\mathcal{F}_{\theta_{c}}$ from scratch on the cleaned dataset.
No detection function, establishing if the training set is poisoned or not, is actually provided by this method (which aims only at cleaning the possibly poisoned dataset).

In~\cite{chen_detecting_2019}, Chen et al.  describe a so-called Activation Clustering (AC) method, that
analyzes the neural network activations  of the last hidden layer (the representation layer), to determine if the training data has been poisoned or not.
The intuition behind this method is that a backdoored model assigns poisoned and benign data to the target class  based on different features, that is, by relying on the triggering pattern for the poisoned samples, and the ground-truth features for the benign ones. This difference is reflected in the representation layer. Therefore, for the target class of the attack, the feature representations of the samples will tend to cluster into two groups, while the representations for the other classes will cluster in one group only.
Based on this intuition, for each subset $\mathcal{D}_{tr,k}$ of $\mathcal{D}_{tr}^{\alpha}$, the defender feeds the images $x_i$ to the model  $\mathcal{F}_{\theta}^{\alpha}$ obtaining the corresponding subset of feature representation vectors or  activations  $f^{-1}_{\theta_{\alpha}}(x_i)$.
Once the activations have been obtained for each training sample,  the subsets
are
clustered separately for each label.
To cluster the activations, the $k$-means algorithm is applied with $k= 2$ (after dimensionality reduction). $k$-means clustering separates the activations into two clusters, regardless of whether
the dataset is poisoned or not. Then, in order to determine
which, if any, of the clusters corresponds to a poisoned subset, one possible approach is to analyze the relative size of the two clusters. A cluster is considered to be poisoned if it contains less than $p$ of data for the $k$ class, that is, $p|\mathcal{D}_{tr,k}|$ samples, where $p\in[0,0.3]$ (the expectation being that poisoned clusters contain no more than a small fraction of class samples, that is $\beta_k \le p$). The corresponding class is detected as the target class.
As a last step, the defender cleans the training dataset, by removing the smallest cluster in the target class, and retraining a new model $\mathcal{F}_{\theta_c}$ from scratch on the cleaned dataset.
As we said, AC can be
applied only when the class poisoning ratio $\beta_k$ is lower than
$p$, ensuring that the poisoned data represents a minority subset
in the target class.
Another method resorting to feature clustering to detect a backdoor attack has been proposed in \cite{soremekun_exposing_2020}.

Even if $k$-means clustering with $k = 2$ can perfectly separate the poisoned data on MNIST and CIFAR-10 when a perceptible triggering pattern is used,
Xiang et al.~\cite{xiang_benchmark_2019} have shown that in many cases, e.g. when the backdoor
pattern is more subtle, the representation vectors of poisoned and benign data can not be separated well in the feature space. This is the case, for instance,  when CIFAR10 is attacked with the single pixel backdoor attack. To improve the results in this case, the authors replace $k$-means clustering with a method based on a Gaussian Mixture Model (GMM), which can also automatically determine the number of clusters.
Under the assumption of subtle (one-pixel) trigger, the authors apply blurring filtering to determine whether a cluster is poisoned or not. After blurring, the samples from the poisoned cluster are assigned to the true class with high probability.
%

A defence working at the training dataset level designed to cope with clean-label backdoor attacks has been proposed in~\cite{peri_deep_2019}.
The defence relies on a so-called deep $k$-Nearest Neighbors ($k\text{-NN}$) defence against feature-collision ~\cite{shafahi_poison_2018} and the  convex polytope \cite{zhu_transferable_2019} attacks mentioned in Section~\ref{sec:cba}. The defence relies on the observation that,  in the representation space, the poisoned samples of a feature collision attack are surrounded by samples having a different label (the target label) (see Fig.~\ref{FIG:feature_collision}). Then,  the authors compare the label of each point $x^{tr}_{i}$ of the training set, with its $k$-nearest neighbors (determined based on the Euclidean distance) in the representation space. If the label of  $x^{tr}$ does  not correspond to  the label of the majority of the $k$ neighbors, $x^{tr}$ is classified as a poisoned sample and removed from the training dataset. Eventually, the network is retrained on the cleaned training dataset to obtain a clean model $\mathcal{F}_{\theta_{c}}$.

As a last example of this class of defences, we mention the work proposed in~\cite{tang_demon_2019}. The defence proposed therein works against source-specific backdoor attacks,  that is, attacks for which the triggering pattern causes a misclassification only when it is added to the images of a specific class (also called targeted contamination attacks). The authors show that this kind of backdoor is more stealthy than source-agnostic backdoors. In this case, in fact, poisoned and benign data can not be easily distinguished by looking at the representation level.
The approach proposed in~\cite{tang_demon_2019} is built upon the {\em universal variation} assumption, according to which the natural variation of the samples of any  uninfected class follows the same distribution of the benign images in the attacked class. For example, in image classification tasks, the natural intra-class variation
of each object (e.g., lighting,
poses, expressions, etc.) has the same distribution across
all labels (this is, for instance, the case of image classification, traffic sign and face recognition tasks).
For such tasks, a DNN model tends to generate a feature representation that can be decomposed
into two parts, one related to the object's identity (e.g. a given individual) and the other depending
on the intra-class variations, randomly drawn from a distribution.
The method described in~\cite{tang_demon_2019} proposes to separate the identity-related features from those associated to the intra-class variations by running
an Expectation-Maximization (EM) algorithm~\cite{chen2012bayesian} across all the representations of the
training samples. Then, if the data distribution of one class is scattered, that class will be likely split into two groups (each group sharing a different identity). If the data distribution is concentrated, the class will be considered as single cluster sharing the same identity. Finally, the defender will judge the class with two groups as an attacked class. 

Other works working at the training dataset level are described below.

Du et al.~\cite{du_robust_2019} have theoretically and empirically proved that applying differential privacy during the training process can efficiently prevent the model from overfitting to the atypical samples. Inspired by this, the authors first add Gaussian noise to the poisoned training dataset, and then utilize it to train an auto-encoder outlier detector. Since poisoned samples are atypical ones, the detector judges one sample to be poisoned if the classification is achieved with less confidence. Finally, Yoshida et al.~\cite{yoshida2020disabling} and Chen et al.~\cite{chen2021pois} share a similar idea for cleaning poisoned data, that is, distilling the clean knowledge from the backdoored model, and further removing poisoned data from the poisoned training dataset by comparing the predictions of the backdoored and distillation models.

\section{Final remarks and research roadmap}
\label{sec.con}
In this work, we have given an overview of backdoor attacks against deep neural networks and possible defences. We started the overview by presenting a unifying framework to cast backdoor attacks in. In doing so, we paid particular attention to define the threat models and the requirements that the attackers and defenders must satisfy under  various settings. Then, we reviewed the main attacks and defences proposed so far, casting them in the general framework outlined previously. This allowed us to critically review the strengths and drawbacks of the various approaches with reference to the application scenarios wherein they are operating.
At the same time, our analysis helps to identify the main open issues still waiting for a solution, thus contributing to outline a roadmap for future research, as described in the rest of this section.


\subsection{Open issues}
\label{sec.reoadmap1}

Notwithstanding the amount of works published so far, there are  several open issues that still remain to be addressed, the most relevant of which are detailed in the following.
\begin{itemize}

  \item {\em More general defences}.
  Existing defences are often tailored solutions that work well only under very specific assumptions about the behavior of the adversary, e.g. on the triggering pattern and its size. In real life applications, however, these assumptions do not necessarily hold. Future research should, then, focus on the development of more general defences, with minimal working assumptions on the attacker's behaviour.
  \item {\em Improving the robustness of backdoors}.  The development of strategies to improve backdoor robustness is another important research line that should occupy the agenda of researchers. Current approaches can resist, up to some extent, to parameter pruning and fine-tuning of final layers, while robustness against retraining of all layers and, more in general, transfer learning, is not at reach of current techniques. Achieving such a robustness is particularly relevant when backdoors are used for benign purposes (see \ref{sec.reoadmap3}). The study of backdoor attacks in the physical domain is another interesting, yet rather unexplored, research direction, (see \cite{xue2021robust} for a preliminary work in this sense), calling for the development of backdoor attacks that can survive  the analog to digital conversion involved by physical domain applications.
      \item {\em Development of an underlying theory}.
           We  ambitiously advocate the need of an underlying theory that can help to solve some of the fundamental problems behind the development of backdoor attacks, like, for instance, the definition of the {\em optimal} triggering pattern (in most of the backdoor attacks proposed so far, the  triggering pattern is  a prescribed signal, arbitrary defined). From the defender's side, a theoretical framework can help the development of more general defences that are effective under a given threat model.
  \item  {\em Video backdoor attacks (and defences)}. Backdoor attacks against video processing networks have attracted significant less interest than attacks working on still images, yet there would be plenty of applications wherein such attacks would be even more relevant than for image-based systems. As a matter of fact, the current literature either focuses on the simple corrupted-label scenario \cite{bhalerao2019luminance}, or it merely applies tools developed for images at the video frame level \cite{zhao_clean-label_2020}. However, for a proper development of video backdoor attacks (and defences), the temporal dimension has to be taken into account, e.g., by designing a triggering pattern that exploits the time dimension of the problem.
\end{itemize}

\subsection{Extension to domains other than computer vision}
\label{sec.reoadmap2}

As mentioned in the introduction, although in this survey we focused on image and video classification,  backdoor attacks and defences
have also been studied in other application domains, e.g., in deep reinforcement learning \cite{kiourti2020trojdrl} and natural language processing \cite{dai2019backdoor}, where, however, the state of the art is less mature.

\subsubsection{Deep reinforcement learning (DRL)}


In 2020, Kiourti et al.~\cite{kiourti2020trojdrl} have presented a backdoor attack against a DRL system.
In this scenario, the backdoored network behaves normally  on untainted states, but works abnormally in some particular states, i.e., the poisoned states, $s_t^*$.
In the non-targeted attack case, the abnormal behavior consists in the agent taking a random action, while for the targeted attack the action taken in correspondence of a poisoned state is a target action chosen by the attacker. The desired abnormal behavior is obtained by poisoning the rewards, assigning a positive reward when the target action is taken in correspondence of $s_t^*$ in the targeted case, or when every action (but the correct one) is taken in the non-targeted case.
According to the result shown in~\cite{kiourti2020trojdrl} a successfull attack is obtained by poisoning a very small percentage of trajectories (states) and rewards.

Some defences to protect a DRL system from backdoor attacks have been also explored in \cite{kiourti2020trojdrl}. It turns out that neither spectral signature~\cite{tran_learning_2015} nor activation clustering~\cite{chen_detecting_2019} can detect the attack because of the small poisoning ratio $\alpha$.
The development of backdoor attacks against DRL system is only at an early stage, and, in particular, the study of effective  backdoor defences is still an open problem.

\subsubsection{Natural language processing (NLP)}

In the NLP domain backdoor attacks and, in particular, defences, are quite advanced.
Starting from \cite{dai2019backdoor}, several works have shown that NLP tools are vulnerable to backdoor attacks. Most of these works implicitly assume that the attack is carried out in a \textit{full control} scenario, where Eve poisons the training dataset in a corrupted-label modality, adding a triggering pattern $\upsilon$, namely, a specific word token, within a benign text sequences, and setting the corresponding label to the target class $t$. The backdoored model will behave as expected on normal text sentences, but will always output $t$ if $\upsilon$ is present in the text string.
%
The first approaches proposed by Kurita et al.~\cite{KuritaMN20} and Wallace et al.~\cite{WallaceZFS21} used noticeable or misspelled words  as trigger $\upsilon$, e.g. `mm', `bb' and `James Bond', that can then be easily detected at test time.  In~\cite{QiYXLS20} and~\cite{QiLCZ0WS20}, a less detectable trigger is used by relying on a proper combination of synonyms and syntaxes.

Two defences ~\cite{WallaceZFS21,azizi2021t} have also been proposed to detect or remove the backdoor from NLP models. Both these methods have serious drawbacks. In~\cite{WallaceZFS21},  the removal of the backdoor reduces the accuracy of the model on benign text, thus not satisfying  the \textit{harmless removal} requirement. The method proposed in~\cite{azizi2021t}, based on the shortcut assumption described in~\cite{wang_neural_2019}, instead, is very time consuming, requiring the computation of a universal perturbation for all possible target classes, which, in NLP applications, can be many. Future work in this area should address the development of clean-labels attacks, and work on more efficient detection and removal methods.

\subsection{Benign uses of backdoors}
\label{sec.reoadmap3}

Before concluding the paper, we pause to mention two possible benign uses of backdoors.

\subsubsection{DNN Watermarking}

Training a   DNN  model is a noticeable  piece of work that requires  significant computational resources
(the training process  may  go on for weeks,
even on powerful machines equipped with several GPUs) and the availability of massive amounts of training data.
For this reason, the demand for methods to protect
the Intellectually Property Rights (IPR) associated to DNNs is rising.
As it happened for media protection \cite{podilchuk2001digital},
watermarking has recently been proposed as a way to protect the
IPRs of DNNs and identify illegitimate usage of DNN models \cite{barni2021dnn}.
In general, the watermark can either be embedded
directly into the weights by modifying the parameters of
one or more layers (static watermarking), or be associated to the behavior
of the network in correspondence to some specific inputs (dynamic
watermarking) \cite{LI2021171}.

The latter approach has immediate
connections with DNN backdooring.
In 2018, Adi et al.~\cite{adi_turning_2018} were the first to propose to black-box watermark a DNN through backdooring. According to~\cite{adi_turning_2018}, the watermark is injected into the DNN during training, by adding a poisoning dataset ($\mathcal{D}_{tr}^{p}$) to the benign training data ($\mathcal{D}_{tr}^{b}$).
The triggering input images in $\mathcal{D}_{tr}^{p}$ play the role of the watermark key.
To verify the ownership, the verifier computes the $ASR$;
if the value is larger than a prescribed threshold the ownership of the DNN is established.

%
In~\cite{adi_turning_2018}, watermark robustness against fine-tuning and transfer learning was evaluated. The results showed that the watermark can be recovered after fine tuning in some cases, while in other cases the accuracy of watermark detection drops dramatically. Transfer learning corresponds to an even more challenging scenario
against which robustness can not be achieved.
Noticeably, poor robustness against transfer learning is a common feature of all DNN watermarking methods developed so far. Improving the robustness of DNN watermarking against network re-use is of primary importance in practical IPR protection applications.
This is linked to the quest for  improving backdoor robustness, already discussed in the previous section.
Moreover,  the use of backdoors for DNN watermarking must be investigated more carefully
in order to understand the capability and the limitations of the backdooring approach in terms of
payload (capacity) and  security, and how it compares  with  static watermarking approaches.

\subsubsection{Trapdoor-enabled adversarial example detection}

DNN models are known to be vulnerable to  adversarial examples, causing  misclassification at testing time \cite{szegedy2013intriguing}. Defense methods developed against adversarial examples work either by designing a system for which adversarial attacks are more difficult to be found (see, for instance, adversarial training~\cite{madry2017towards} and defensive distillation~\cite{papernot2016distillation}), or by trying to detect the adversarial inputs at testing time (e.g., by feature squeezing, or input pre-processing \cite{liao2018defense}).

Recently,  Shan et al.~\cite{shan_gotta_2019}  have proposed to exploit backdoor attacks to protect DNN models against adversarial examples, by implementing a so-called trapdoor honeypot.
A trapdoor honeypot is similar to a backdoor in that it causes a misclassification error in the presence of a specific, minimum energy, triggering pattern. When building an adversarial example, the attacker will likely, and inadvertently, exploit the weakness introduced within the DNN by the backdoor and come out with an adversarial perturbation which is very close to the triggering pattern purposely introduced by the defender at training time. In this way, the defender may recognize that an adversarial attack is ongoing and react accordingly.

More specifically, given a to-be-protected class $t$, the defender trains a backdoored model $\mathcal{F}_{\theta_{\alpha}^{*}}$ such that  $\mathcal{F}_{\theta_{\alpha}^{*}}(x+\upsilon)=t \neq \mathcal{F}_{\theta_{\alpha}^{*}}(x)$, where  $\upsilon$  is a low-energy triggering pattern, called loss-minimizing
trapdoor, designed in such a way
to minimize the loss for the target label.
The presence of an adversarial input can then be detected by looking for the presence of the pattern $\upsilon$ within the input sample, trusting that the algorithm used to construct the adversarial perturbation will exploit the existence of a low-energy pattern $\upsilon$ capable of inducing a misclassification error in favour of class $t$.
Based on the results shown in ~\cite{shan_gotta_2019}, the trapdoor-enabled defence achieves high accuracy against many state-of-art targeted adversarial examples attacks.

Such defense works only  against targeted attacks, and  trapdoor honeypots  against  non-targeted adversarial example have still to be developed.
Moreover, how to extend the idea of trapdoor honeypots to defend against black-box adversarial examples, that do not adopt a low-energy pattern, is an open issue deserving further attention.

\section{Acknowledgment}
This work has been partially supported by the Italian Ministry of University and Research under the PREMIER project, and by the China Scholarship Council (CSC), file No.201908130181.

\bibliographystyle{IEEEtran}
\bibliography{backdoor}

\end{document}